\documentclass[10pt]{article}

\newcommand{\beq} {\begin{equation}}
\newcommand{\enq} {\end{equation}}
\newcommand{\ber} {\begin {eqnarray}}
\newcommand{\enr} {\end {eqnarray}}
\newcommand{\eq} {equation}
\newcommand{\eqn} {equation }
\newcommand{\eqs} {equations }
\newcommand{\ens} {equations}

\newcommand {\er}[1] {equation (\ref{#1}) }
\newcommand {\ern}[1] {equation (\ref{#1})}
\newcommand {\ers}[1] {equations (\ref{#1})}
\newcommand {\Er}[1] {Equation (\ref{#1}) }

\newcommand*{\ds}{\displaystyle}
\providecommand*{\dfrac}[2]{\ds\frac{#1}{#2}}
\newcommand*{\dR}{{d}}

\newcommand*{\Dt}[1]{\dfrac{d{#1}}{dt}}

\newcommand*{\od}[2]{\dfrac{{\dR}{#1}}{{\dR}{#2}}}

\title{Noether Currents for Eulerian Variational Principles in Non Barotropic Magnetohydrodynamics and Topological Conservations Laws}
\author{Asher Yahalom$^{a,b}$ and Hong Qin$^b$ \\
$^a$ Ariel University, Kiryat Hamada POB 3, Ariel 40700, Israel\\
$^b$ Princeton University, Princeton, New Jersey 08543, USA\\
e-mails: asya@ariel.ac.il, hongqin@princeton.edu}

\begin{document}
\maketitle

\begin {abstract}
We derive a Noether current for the Eulerian variational principle of ideal non-barotropic magnetohydrodynamics (MHD). It was shown  previously that ideal non-barotropic MHD is mathematically equivalent to a five function field theory with an induced geometrical structure  in the case that field lines cover surfaces and this theory can be described using a variational principle. Here we use various symmetries of the flow to derive topological constants of motion through the derived Noether current and discuss their implication for non-barotropic MHD.

\vspace{0.3cm}
\noindent Keywords: Magnetohydrodynamics, Variational principles, Topological Constants of Motion

\vspace{0.3cm}
\noindent PACS number(s): 52.30.Cv, 02.30.Xx
\end {abstract}

\section {Introduction}

Variational principles for MHD were introduced by
previous authors both in Lagrangian and Eulerian form.
Vladimirov and Moffatt \cite{Moffatt} in a series of papers have discussed an Eulerian
variational principle for incompressible MHD.
However, their variational principle contained three more
functions in addition to the seven variables which appear in the
standard equations of incompressible MHD which are the magnetic
field $\vec B$ the velocity field $\vec v$ and the pressure $P$.
 Yahalom \& Lynden-Bell \cite{YaLy} obtained an {\bf Eulerian} Lagrangian principle for barotropic MHD
which will depend on only six functions. The variational
derivative of this Lagrangian produced all the equations
needed to describe barotropic MHD without any
additional constraints. Yahalom \cite{Yah} have shown that for the barotropic case four functions will
suffice. Moreover, it was shown that the cuts of some of those functions \cite{Yah2}
are topological local conserved quantities.

Previous work was concerned only with barotropic magnetohydrodynamics. Variational principles of non
barotropic magnetohydrodynamics can be found in the work of
Bekenstein \& Oron \cite{Bekenstien} in terms of 15 functions and
V.A. Kats \cite{Kats} in terms of 20 functions. Morrison \cite{Morrison}
has suggested a Hamiltonian approach but this also depends on 8 canonical variables (see table 2 \cite{Morrison}).

It was shown that this number can be somewhat reduced. In \cite{nonBarotropic,nonBarotropic2} it was demonstrated that only five functions will
suffice to describe non barotropic magnetohydrodynamics, and that the
reduced lagrangian has a distinct geometrical structure including an induced metric.

The theorem of Noether dictates that for every continuous symmetry group of an Action the system must possess a conservation law.
For example time translation symmetry results in the conservation of energy, while spatial translation symmetry results in the conservation of linear momentum and rotation symmetry in the conservation of angular momentum to list some well known examples. But sometimes the conservation law is discovered without reference to the Noether theorem by using the equations of the system. In that case one is tempted to inquire what is
the hidden symmetry associated with this conservation law and what is the simplest way to represent it.

The concept of metage as a label for fluid elements along a vortex line in ideal fluids was first introduced by
Lynden-Bell \& Katz \cite{LynanKatz}. A translation group of this label was found to be connected to the conservation
of Moffat's \cite{Moffatt} helicity by Yahalom \cite{Yahalomhel} using a {\bf Lagrangian} variational principle. The concept of metage was later generalized by Yahalom \& Lynden-Bell \cite{YaLy} for barotropic MHD, but now as a label for fluid elements along magnetic field lines which are comoving with the flow in the case of ideal MHD. Yahalom \& Lynden-Bell \cite{YaLy} has also shown that the translation group of the magnetic metage is connected to Woltjer \cite{Woltjer1,Woltjer2} conservation of cross helicity for barotropic MHD. Recently the concept of metage was generalized also for non barotropic MHD in which magnetic field lines lie on entropy surfaces \cite{simpvarYah}. This was later generalized by dropping the entropy condition on magnetic field lines \cite{metra}. In those papers the metage translation symmetry group was used to generate a non-barotropic cross helicity generalization using a {\bf Lagrangian} variational principle.

Cross Helicity was first described by Woltjer \cite{Woltjer1,Woltjer2} and is give by:
\begin{equation} \label{GrindEQ__22_}
H_{C} \equiv \int  \vec{B}\cdot \vec{v}d^{3} x,
\end{equation}
in which the integral is taken over the entire flow domain. $H_{C}$ is conserved for barotropic or incompressible MHD and
is given a topological interpretation in terms of the knottiness of magnetic and flow field lines.

Both conservation laws for the helicity in the fluid dynamics case and the barotropic MHD case were
shown to originate from a relabelling symmetry through the Noether theorem \cite{YaLy,Yahalomhel,Padhye1,Padhye2}.
Webb et al. \cite{Webb2} have generalized the idea of relabelling symmetry to non-barotropic MHD and derived their generalized cross helicity conservation law by using Noether's theorem but without using the simple representation which is connected with the metage variable. The conservation law deduction involves a divergence symmetry of the action. These conservation laws were written as Eulerian conservation laws of the form $D_t+\vec \nabla \cdot \vec F = 0$ where D is the conserved density and F is the conserved flux. Webb et al. \cite{Webb4} discuss the cross helicity conservation law for non-barotropic MHD in a multi-symplectic formulation of MHD. Webb et al. \cite{Webb1,Webb2} emphasize that the generalized cross helicity conservation law, in MHD and the generalized helicity conservation law in non-barotropic fluids are non-local in the sense that they depend on the auxiliary nonlocal variable $\sigma$, which depends on the Lagrangian time integral of the temperature $T(x, t)$. Notice that a potential vorticity conservation equation for non-barotropic MHD is derived by Webb, G. M. and Mace, R.L. \cite{Webb5} by using Noether's second theorem.

Recently the non-barotropic cross helicity was generalized using additional label translation symmetry groups ($\chi$ and $\eta$ translations) \cite{chiettrans}, this led to additional topological conservation laws the $\chi$ and $\eta$ cross helicities.

Previous analysis depended on {\bf Lagrangian} variational principles and their Noether currents. Here we introduce a novel approach
based on an {\bf Eulerian} variational principle. We derive the Noether current of the Eulerian variational principle and show how this can be
used to derive topological conservation laws using label symmetries.

The plan of this paper is as follows: First we introduce the
standard notations and equations of non-barotropic
magnetohydrodynamics. Next we introduce the Eulerian variational principle suitable for the non-barotropic case.
This is followed by a derivations of the Noether Current and finally we use the Noether current to obtain the generalized non-
barotropic cross helicities. Implication for non-barotropic MHD dynamics of the topological conservation laws are discussed.

\section{Standard formulation of ideal non-barotropic magnetohydrodynamics}

The standard set of \eqs solved for non-barotropic magnetohydrodynamics are given below:

\beq
\frac{\partial{\vec B}}{\partial t} = \vec \nabla \times (\vec v \times \vec B)
\label{Beq}
\enq
\beq
\vec \nabla \cdot \vec B =0
\label{Bcon}
\enq
\beq
\frac{\partial{\rho}}{\partial t} + \vec \nabla \cdot (\rho \vec v ) = 0
\label{masscon}
\enq
\beq
\rho \frac{d \vec v}{d t}=
\rho (\frac{\partial \vec v}{\partial t}+(\vec v \cdot \vec \nabla)\vec v)  = -\vec \nabla p (\rho,s) +
\frac{(\vec \nabla \times \vec B) \times \vec B}{4 \pi}
\label{Euler}
\enq
\beq
 \frac{d s}{d t}=0
\label{Ent}
\enq
The following notations are utilized: $\frac{\partial}{\partial t}$ is the temporal derivative,
$\frac{d}{d t}$ is the temporal material derivative and $\vec \nabla$ has its
standard meaning in vector calculus.  $\rho$ is the fluid density and $s$ is the specific entropy. Finally $p (\rho,s)$ is the pressure which
depends on the density and entropy (the non-barotropic case). The justification for those \eqs
and the conditions under which they apply can be
found in standard books on magnetohydrodynamics (see for example \cite{Sturrock}).
The number of independent variables for which one needs to solve is eight
($\vec v,\vec B,\rho,s$) and the number of \eqs (\ref{Beq},\ref{masscon},\ref{Euler},\ref{Ent}) is also eight.
Notice that \ern{Bcon} is a condition on the initial $\vec B$ field and is satisfied automatically for
any other time due to \ern{Beq}. We will find it useful to introduce the following thermodynamic equations for later use:
\ber
d \varepsilon &=& T ds - p d \frac{1}{\rho} = T ds + \frac{p}{\rho^2} d \rho
\nonumber \\
& & \frac{\partial \varepsilon}{\partial s}_{\rho} = T, \qquad \frac{\partial \varepsilon}{\partial \rho}_{s} = \frac{p}{\rho^2}
\nonumber \\
w &=& \varepsilon + \frac{p}{\rho}= \varepsilon + \frac{\partial \varepsilon}{\partial \rho} \rho = \frac{\partial (\rho \varepsilon)}{\partial \rho}
\nonumber \\
dw &=& d\varepsilon + d(\frac{p}{\rho}) = T ds +  \frac{1}{\rho} dp
\label{thermodyn}
\enr
in the above: $\varepsilon$ is the specific internal energy, $T$ is the temperature and $w$ is the specific enthalpy. A special case of equation of state is the polytropic equation of state \cite{bt}:
\beq
p =K \rho^{\gamma}
\label{plytrop}
\enq
$K$ and $\gamma$ may depend on the specific entropy $s$. Hence:
\beq
\frac{\partial \varepsilon}{\partial \rho} = K \rho^{\gamma-2} \Rightarrow \varepsilon = \frac{K}{\gamma-1} \rho^{\gamma-1}
 =  \frac{p}{\rho(\gamma-1)}  \Rightarrow \rho \varepsilon = \frac{p}{\gamma-1}
\label{plytrop2}
\enq
the last identity is up to a function dependent on $s$.

\section{Variational principle of non-barotropic magnetohydrodynamics}

In the following section we will generalize the approach of \cite{YaLy} for the non-barotropic case \cite{nonBarotropic,nonBarotropic2}.
Consider the action:

\ber A & \equiv & \int {\cal L} d^3 x dt,
\nonumber \\
{\cal L} & \equiv & {\cal L}_1 + {\cal L}_2,
\nonumber \\
{\cal L}_1 & \equiv & \rho (\frac{1}{2} \vec v^2 - \varepsilon (\rho,s)) +  \frac{\vec B^2}{8 \pi},
\nonumber \\
{\cal L}_2 & \equiv & \nu [\frac{\partial{\rho}}{\partial t} + \vec \nabla \cdot (\rho \vec v )]
- \rho \alpha \frac{d \chi}{dt} - \rho \beta \frac{d \eta}{dt} - \rho \sigma \frac{d s}{dt}
\nonumber \\
 &-& \frac{\vec B}{4 \pi} \cdot \vec \nabla \chi \times \vec \nabla \eta.
\label{Lagactionsimp}
\enr
In the specific case of a polytropic equation of state we have according to \ern{plytrop2}:
\beq
{\cal L}_1 = \frac{1}{2} \rho  \vec v^2 -  \frac{p}{\gamma-1} +  \frac{\vec B^2}{8 \pi}.
\label{Lagactionsimp2}
\enq
Obviously $\nu,\alpha,\beta,\sigma$ are Lagrange multipliers which were inserted in such a
way that the variational principle will yield the following \ens:
\ber
& & \frac{\partial{\rho}}{\partial t} + \vec \nabla \cdot (\rho \vec v ) = 0,
\nonumber \\
& & \rho \frac{d \chi}{dt} = 0, \rho \frac{d \eta}{dt} = 0, \rho \frac{d s}{dt} = 0.
\label{lagmul}
\enr
It {\bf is not} assumed that $\nu,\alpha,\beta,\sigma$  are single valued.
Provided $\rho$ is not null those are just the continuity \ern{masscon}, entropy conservation
 and the conditions that Sakurai's functions are comoving.
Taking the variational derivative with respect to $\vec B$ we see that:
\beq
\vec B = \hat {\vec B} \equiv \vec \nabla \chi \times \vec \nabla \eta.
\label{Bsakurai2}
\enq
Hence $\vec B$ is in Sakurai's form \cite{Sakurai} and satisfies \ern{Bcon}.
It can be easily shown that provided that $\vec B$ is in the form given in \ern{Bsakurai2},
and \ers{lagmul} are satisfied, then also \ern{Beq} is satisfied. We notice that the specific form of the magnetic field given in \ern{Bsakurai2} appear under
different names in the literature. The functions $\chi$ and $\eta$ are sometimes denoted "Euler potentials", "Clebsch variables" and also
"flux representation functions" \cite{Con}. \Er{Bsakurai2} imply that the magnetic field lines lie on surfaces, the lines may be surface filling
but not volume filling.

For the time being we have showed that all the equations of non-barotropic magnetohydrodynamics can be obtained
from the above variational principle except Euler's equations. We will now
show that Euler's equations can be derived from the above variational principle
as well. Let us take an arbitrary variational derivative of the above action with
respect to $\vec v$, this will result in:

\ber
\delta_{\vec v} A &=& \hspace{-0.2cm} \int dt \{ \int d^3 x dt \rho \delta \vec v \cdot
[\vec v - \vec \nabla \nu - \alpha \vec \nabla \chi - \beta \vec \nabla \eta
\nonumber \\
&-& \sigma \vec \nabla s]
+ \oint d \vec S \cdot \delta \vec v \rho \nu+  \int d \vec \Sigma \cdot \delta \vec v \rho [\nu]\}.
\label{delActionv}
\enr
The integral $\oint d \vec S \cdot \delta \vec v \rho \nu$ vanishes in many physical scenarios.
In the case of astrophysical flows this integral will vanish since $\rho=0$ on the flow
boundary, in the case of a fluid contained
in a vessel no flux boundary conditions $\delta \vec v \cdot \hat n =0$ are induced
($\hat n$ is a unit vector normal to the boundary). The surface integral $\int d \vec \Sigma$
 on the cut of $\nu$ vanishes in the case that $\nu$ is single valued and $[\nu]=0$ .
In the case that $\nu$ is not single valued only a Kutta type velocity perturbation \cite{YahPinhasKop} in which
the velocity perturbation is parallel to the cut will cause the cut integral to vanish.

Provided that the surface integrals do vanish and that $\delta_{\vec v} A =0$ for an arbitrary
velocity perturbation we see that $\vec v$ must have the following form:

\beq
\vec v = \hat {\vec v} \equiv \vec \nabla \nu + \alpha \vec \nabla \chi + \beta \vec \nabla \eta + \sigma \vec \nabla s.
\label{vform}
\enq
The above equation is reminiscent of Clebsch representation in non magnetic fluids. A similar expression
was obtained by Morrison \cite{Morrison} using an Hamiltonian formalism but in which the $s$ terms is
replaced by $\psi$ which is conjugate to $s$. Let us now take the variational derivative with respect to the density $\rho$ we obtain:

\ber
\delta_{\rho} A & = & \int d^3 x dt \delta \rho
[\frac{1}{2} \vec v^2 - w  - \frac{\partial{\nu}}{\partial t} -  \vec v \cdot \vec \nabla \nu]
\nonumber \\
 & + & \int dt \oint d \vec S \cdot \vec v \delta \rho  \nu +
  \int dt \int d \vec \Sigma \cdot \vec v \delta \rho  [\nu]
\nonumber \\
  &+& \int d^3 x \nu \delta \rho |^{t_1}_{t_0}.
\label{delActionrho}
\enr
In which $ w= \frac{\partial (\varepsilon \rho)}{\partial \rho}$ is the specific enthalpy.
Hence provided that $\oint d \vec S \cdot \vec v \delta \rho  \nu$ vanishes on the boundary of the domain
and $ \int d \vec \Sigma \cdot \vec v \delta \rho  [\nu]$ vanishes on the cut of $\nu$
in the case that $\nu$ is not single valued\footnote{Which entails either a Kutta type
condition for the velocity or a vanishing density perturbation on the cut.}
and in initial and final times the following \eqn must be satisfied:

\beq
\frac{d \nu}{d t} = \frac{1}{2} \vec v^2 - w, \qquad
\label{nueq}
\enq
Finally we have to calculate the variation with respect to both $\chi$ and $\eta$
this will lead us to the following results:

\ber
\delta_{\chi} A \hspace{-0.4cm} & = & \hspace{-0.4cm} \int d^3 x dt \delta \chi
[\frac{\partial{(\rho \alpha)}}{\partial t} +  \vec \nabla \cdot (\rho \alpha \vec v)-
\vec \nabla \eta \cdot \vec J]
 \nonumber \\
&+& \int dt \oint d \vec S \cdot [\frac{\vec B}{4 \pi} \times \vec \nabla \eta - \vec v \rho \alpha]\delta \chi
 \nonumber \\
 & + & \int dt \int d \vec \Sigma \cdot [\frac{\vec B}{4 \pi} \times \vec \nabla \eta - \vec v \rho \alpha][\delta \chi]
 \nonumber \\
 &-& \int d^3 x \rho \alpha \delta \chi |^{t_1}_{t_0},
\label{delActionchi}
\enr

\ber
\delta_{\eta} A \hspace{-0.4cm} & = & \hspace{-0.4cm} \int d^3 x dt \delta \eta
[\frac{\partial{(\rho \beta)}}{\partial t} +  \vec \nabla \cdot (\rho \beta \vec v)+
\vec \nabla \chi \cdot \vec J]
\nonumber \\
&+& \int dt \oint d \vec S \cdot [\vec \nabla \chi \times \frac{\vec B}{4 \pi} - \vec v \rho \beta]\delta \eta
\nonumber \\
 & + &  \int dt \int d \vec \Sigma \cdot [\vec \nabla \chi \times \frac{\vec B}{4 \pi} - \vec v \rho \beta][\delta \eta]
\nonumber \\
 &-& \int d^3 x \rho \beta \delta \eta |^{t_1}_{t_0}.
\label{delActioneta}
\enr
Provided that the correct temporal and boundary conditions are met with
respect to the variations $\delta \chi$ and $\delta \eta$ on the domain boundary and
on the cuts in the case that some (or all) of the relevant functions are non single valued.
we obtain the following set of equations:

\beq
\frac{d \alpha}{dt} = \frac{\vec \nabla \eta \cdot \vec J}{\rho}, \qquad
\frac{d \beta}{dt} = -\frac{\vec \nabla \chi \cdot \vec J}{\rho},
\label{albetaeq}
\enq
in which the continuity \ern{masscon} was taken into account. By correct temporal conditions we
mean that both $\delta \eta$ and $\delta \chi$ vanish at initial and final times. As for boundary
conditions which are sufficient to make the boundary term vanish on can consider the case that
the boundary is at infinity and both $\vec B$ and $\rho$ vanish. Another possibility is that the boundary is
impermeable and perfectly conducting. A sufficient condition for the integral over the "cuts" to vanish
is to use variations $\delta \eta$ and $\delta \chi$ which are single valued. It can be shown that
$\chi$ can always be taken to be single valued, hence taking $\delta \chi$ to be single valued is no
restriction at all. In some topologies $\eta$ is not single valued and in those cases a single valued
restriction on $\delta \eta$ is sufficient to make the cut term null.

Finally we take a variational derivative with respect to the entropy $s$:

\ber
\delta_{s} A \hspace{-0.4cm} & = & \hspace{-0.4cm} \int d^3 x dt \delta s
[\frac{\partial{(\rho \sigma)}}{\partial t} +  \vec \nabla \cdot (\rho \sigma \vec v)- \rho T]
 \nonumber \\
&+& \int dt \oint d \vec S \cdot \rho \sigma \vec v  \delta s
 -  \int d^3 x \rho \sigma \delta s |^{t_1}_{t_0},
\label{delActions}
\enr
in which the temperature is $T=\frac{\partial \varepsilon}{\partial s}$. We notice that according
to \ern{vform} $\sigma$ is single valued and hence no cuts are needed. Taking into account the continuity
\ern{masscon} we obtain for locations in which the density $\rho$ is not null the result:

\beq
\frac{d \sigma}{dt} =T,
\label{sigmaeq}
\enq
provided that $\delta_{s} A$ vanished for an arbitrary $\delta s$.

\subsection{Euler's equations}

We shall now show that a velocity field given by \ern{vform}, such that the
\eqs for $\alpha, \beta, \chi, \eta, \nu, \sigma, s$ satisfy the corresponding equations
(\ref{lagmul},\ref{nueq},\ref{albetaeq},\ref{sigmaeq}) must satisfy Euler's equations.
Let us calculate the material derivative of $\vec v$:

\ber
\frac{d\vec v}{dt} &=& \frac{d\vec \nabla \nu}{dt}  + \frac{d\alpha}{dt} \vec \nabla \chi +
 \alpha \frac{d\vec \nabla \chi}{dt}  +
\frac{d\beta}{dt} \vec \nabla \eta + \beta \frac{d\vec \nabla \eta}{dt}
\nonumber \\
&+& \frac{d\sigma}{dt} \vec \nabla s + \sigma \frac{d\vec \nabla s}{dt}
\label{dvform}
\enr
It can be easily shown that:

\ber
\frac{d\vec \nabla \nu}{dt} & = & \vec \nabla \frac{d \nu}{dt}- \vec \nabla v_k \frac{\partial \nu}{\partial x_k}
 = \vec \nabla (\frac{1}{2} \vec v^2 - w)- \vec \nabla v_k \frac{\partial \nu}{\partial x_k},
 \nonumber \\
 \frac{d\vec \nabla \eta}{dt} & = & \vec \nabla \frac{d \eta}{dt}- \vec \nabla v_k \frac{\partial \eta}{\partial x_k}
 = - \vec \nabla v_k \frac{\partial \eta}{\partial x_k},
 \nonumber \\
 \frac{d\vec \nabla \chi}{dt} & = & \vec \nabla \frac{d \chi}{dt}- \vec \nabla v_k \frac{\partial \chi}{\partial x_k}
 = - \vec \nabla v_k \frac{\partial \chi}{\partial x_k},
  \nonumber \\
 \frac{d\vec \nabla s}{dt} & = & \vec \nabla \frac{d s}{dt}- \vec \nabla v_k \frac{\partial s}{\partial x_k}
 = - \vec \nabla v_k \frac{\partial s}{\partial x_k}.
 \label{dnabla}
\enr
In which $x_k$ is a Cartesian coordinate and a summation convention is assumed. Inserting the result from equations (\ref{dnabla},\ref{lagmul})
into \ern{dvform} yields:

\ber
\frac{d\vec v}{dt} &=& - \vec \nabla v_k (\frac{\partial \nu}{\partial x_k} + \alpha \frac{\partial \chi}{\partial x_k} +
\beta \frac{\partial \eta}{\partial x_k} + \sigma \frac{\partial s}{\partial x_k})
 \nonumber \\
&+& \vec \nabla (\frac{1}{2} \vec v^2 - w)+ T \vec \nabla s
 \nonumber \\
&+& \frac{1}{\rho} ((\vec \nabla \eta \cdot \vec J)\vec \nabla \chi
- (\vec \nabla \chi \cdot \vec J)\vec \nabla \eta)
 \nonumber \\
&=& - \vec \nabla v_k v_k + \vec \nabla (\frac{1}{2} \vec v^2 - w) + T \vec \nabla s
 \nonumber \\
 &+& \frac{1}{\rho} \vec J \times (\vec \nabla \chi \times  \vec \nabla \eta)
 \nonumber \\
&=& - \frac{\vec \nabla p}{\rho} + \frac{1}{\rho} \vec J \times \vec B.
\label{dvform2}
\enr
In which we have used both \ern{vform} and \ern{Bsakurai2} in the above derivation. This of course
proves that the non-barotropic Euler equations can be derived from the action given in \er{Lagactionsimp} and hence
all the equations of non-barotropic magnetohydrodynamics can be derived from the above action
without restricting the variations in any way except on the relevant boundaries and cuts.

\subsection{Local non-barotropic cross helicity}

The function $\nu$, whose material derivative is given in (\ref{nueq}),
can be multiple valued because only its gradient appears in the velocity (\ref{vform}). However, the discontinuity, $[\nu ]$, of $\nu $ is conserved,
\begin{equation}
\Dt{[\nu]} =0
\label{GrindEQ__37_}
\end{equation}
since the terms on the right-hand side of (\ref{nueq})
describe  physical quantities and hence are single valued. A similar equation also holds for barotropic fluid dynamics and barotropic MHD \cite{YaLy,Yah,Yah2}.

We now substitute the expressions for ${\vec B}$ and ${\vec v}$ given by (\ref{Bsakurai2}) and (\ref{vform}) respectively into the formula $H_{C} \equiv \int  \vec{B}\cdot \vec{v}\dR^{3} x$ for the cross helicity (see (\ref{GrindEQ__22_})) to obtain
\begin{equation} \label{GrindEQ__22d_}
H_{C} = \int \dR \Phi [\nu] + \int \dR \Phi \oint \sigma d s,
\end{equation}
where the closed line integral taken along a magnetic field line. Furthermore, $\dR\Phi =\vec{B}\cdot d \vec{S}=(\vec{\nabla }\chi \times \vec{\nabla }\eta) \cdot d \vec{S}=\dR\chi  \dR\eta $ is a magnetic flux element which is co-moving as governed by (\ref{Beq}) and $\dR\vec{S}$ is an infinitesimal area element. Although the cross helicity is not conserved for non-barotropic flows, inspection of the right-hand side of (\ref{GrindEQ__22d_}) reveals that it is made of a sum of two terms. One term is conserved, as both $\dR\Phi$ and $[\nu]$ are co-moving,  and the other is not. This suggests the following definition for the non-barotropic cross helicity
\begin{equation}
H_{CNB}  \equiv  \int \dR \Phi  [\nu]  =  H_{C}  -   \int \dR \Phi \oint \sigma\dR s .
\label{GrindEQ__22e_}
\end{equation}
It can be written in the more conventional form:
\begin{equation} \label{GrindEQ__22e1_}
H_{CNB}  =  \int  \vec{B} \cdot \vec v_t   \dR^{3} x
\end{equation}
in which the topological velocity field is defined as
\begin{equation}
 \vec v_t  =  \vec v  -  \sigma \vec \nabla s .
 \label{vt}
\end{equation}
It should be noted that $H_{CNB}$ is conserved even for an MHD not satisfying the Sakurai topological constraint given in (\ref{Bsakurai2}), provided that we have a field $\sigma$ satisfying the equation $\Dt{\sigma} = T$.
  This can be verified by direct derivation using only the equation of motion and the sigma equation. Thus the non-barotropic cross helicity conservation law,
\begin{equation}
\od{H_{CNB}}{t}  =  0 ,
\label{HCNBcon}
\end{equation}
is more general than the variational principle described by (\ref{Lagactionsimp8}) as follows from a direct computation using
(\ref{Beq}) and (\ref{masscon})--(\ref{Ent}).
Also note that, for a constant specific entropy $s$, we obtain $H_{CNB}=H_{C}$ and the non-barotropic cross helicity reduces to the standard barotropic cross helicity. The local form of \ern{HCNBcon} describing the evolution of $H_{CNB}$ per unit volume was described by \cite{Webb1,Webb2}. To conclude we introduce also a {\bf local} topological conservation law in the spirit of \cite{Yah2} which is the non-barotropic cross helicity per unit of magnetic flux. This quantity which is equal to the discontinuity, $[\nu]$, of $\nu$  is conserved and can be written as a sum of the barotropic cross helicity per unit flux and the closed line integral of $s \dR \sigma$ along a magnetic field line, namely:
\begin{equation} \label{loc_}
[\nu]= \od{H_{CNB}}{\Phi} = \od{H_{C}}{\Phi} + \oint  s \dR \sigma.
\end{equation}

\subsection{Simplified action}

The reader of this paper might argue here that the paper is misleading. The author has declared
that he is going to present a simplified action for non-barotropic magnetohydrodynamics instead he
 added six more functions $\alpha,\beta,\chi,\-\eta,\nu,\sigma$ to the standard set $\vec B,\vec v,\rho,s$.
In the following I will show that this is not so and the action given in \ern{Lagactionsimp} in
a form suitable for a pedagogic presentation can indeed be simplified. It is easy to show
that the Lagrangian density appearing in \ern{Lagactionsimp} can be written in the form:

\ber
{\cal L} & = & -\rho [\frac{\partial{\nu}}{\partial t} + \alpha \frac{\partial{\chi}}{\partial t}
+ \beta \frac{\partial{\eta}}{\partial t}+ \sigma \frac{\partial{s}}{\partial t}+\varepsilon (\rho,s)]
\nonumber \\
&+&
\frac{1}{2}\rho [(\vec v-\hat{\vec v})^2-(\hat{\vec v})^2]
\nonumber \\
& + &   \frac{1}{8 \pi} [(\vec B-\hat{\vec B})^2-(\hat{\vec B})^2]+
\frac{\partial{(\nu \rho)}}{\partial t} + \vec \nabla \cdot (\nu \rho \vec v )
\label{Lagactionsimp4}
\enr
In which $\hat{\vec v}$ is a shorthand notation for $\vec \nabla \nu + \alpha \vec \nabla \chi +
 \beta \vec \nabla \eta +  \sigma \vec \nabla s $ (see \ern{vform}) and $\hat{\vec B}$ is a shorthand notation for
 $\vec \nabla \chi \times \vec \nabla \eta$ (see \ern{Bsakurai2}). Thus ${\cal L}$ has four contributions:

\ber
  {\cal L}  &  = &  \hat {\cal L} + {\cal L}_{\vec v}+ {\cal L}_{\vec B}+{\cal L}_{boundary},
\nonumber \\
\hat {\cal L}   \hspace{1.2 cm} &\hspace{-2.5 cm} \equiv &  \hspace{-1.5 cm} -\rho [\frac{\partial{\nu}}{\partial t}
+ \alpha \frac{\partial{\chi}}{\partial t}
+ \beta \frac{\partial{\eta}}{\partial t}+ \sigma \frac{\partial{s}}{\partial t}+\varepsilon (\rho,s)
\nonumber \\
&+& \frac{1}{2} (\vec \nabla \nu + \alpha \vec \nabla \chi +  \beta \vec \nabla \eta +  \sigma \vec \nabla s )^2 ]
\nonumber \\
&-&\frac{1}{8 \pi}(\vec \nabla \chi \times \vec \nabla \eta)^2
\nonumber \\
{\cal L}_{\vec v} &\equiv & \frac{1}{2}\rho (\vec v-\hat{\vec v})^2,
\nonumber \\
{\cal L}_{\vec B} &\equiv & \frac{1}{8 \pi} (\vec B-\hat{\vec B})^2,
\nonumber \\
{\cal L}_{boundary} &\equiv & \frac{\partial{(\nu \rho)}}{\partial t} + \vec \nabla \cdot (\nu \rho \vec v ).
\label{Lagactionsimp5}
\enr
The only term containing $\vec v$ is\footnote{${\cal L}_{boundary}$ also depends on
$\vec v$ but being a boundary term in space and time it does not contribute to the derived equations}
 ${\cal L}_{\vec v}$, it can easily be seen that
this term will lead, after we nullify the variational derivative with respect to $\vec v$,
to \ern{vform} but will otherwise
have no contribution to other variational derivatives. Similarly the only term containing $\vec B$
is ${\cal L}_{\vec B}$ and it can easily be seen that
this term will lead, after we nullify the variational derivative, to \ern{Bsakurai2} but will
have no contribution to other variational derivatives. Also notice that the term ${\cal L}_{boundary}$
contains only complete partial derivatives and thus can not contribute to the equations although
it can change the boundary conditions. Hence we see that \ers{lagmul}, \ern{nueq}, \ers{albetaeq} and \er{sigmaeq}
can be derived using the Lagrangian density:
\ber
& & \hat {\cal L}[\alpha,\beta,\chi,\eta,\nu,\rho,\sigma,s] = -\rho [\frac{\partial{\nu}}{\partial t} + \alpha \frac{\partial{\chi}}{\partial t}
+ \beta \frac{\partial{\eta}}{\partial t}
\nonumber \\
&+& \sigma \frac{\partial{s}}{\partial t}
 +\  \varepsilon (\rho,s) + \frac{1}{2} (\vec \nabla \nu + \alpha \vec \nabla \chi +  \beta \vec \nabla \eta +  \sigma \vec \nabla s )^2 ]
\nonumber \\
&-&\frac{1}{8 \pi}(\vec \nabla \chi \times \vec \nabla \eta)^2
\label{Lagactionsimp6}
\enr
in which $\hat{\vec v}$ replaces $\vec v$ and $\hat{\vec B}$ replaces $\vec B$ in the rele\-vant equations.
Furthermore, after integrating the ei\-ght \eqs (\ref{lagmul},\ref{nueq},\ref{albetaeq},\ref{sigmaeq})
 we can insert the potentials $\alpha,\beta,\chi,\eta,\nu,\sigma,s$
into \ers{vform} and (\ref{Bsakurai2}) to obtain the physical quantities $\vec v$ and $\vec B$.
Hence, the general non-barotropic magnetohydrodynamic problem is reduced from eight equations
(\ref{Beq},\ref{masscon},\ref{Euler},\ref{Ent}) and the additional constraint (\ref{Bcon})
to a problem of eight first order (in the temporal derivative) unconstrained equations.
Moreover, the entire set of equations can be derived from the Lagrangian density $\hat {\cal L}$.

\subsection{Further Simplification}

\subsubsection{Elimination of Variables}

 Let us now look at the three last three equations of (\ref{lagmul}) \cite{nonBarotropic,nonBarotropic2}. Those describe three comoving quantities
 which can be written in terms of the generalized Clebsch form given in \ern{vform} as follows:

\ber
& &  \frac{\partial \chi}{\partial t} + (\vec \nabla \nu + \alpha \vec \nabla \chi + \beta \vec \nabla \eta + \sigma \vec \nabla s)
\cdot \vec \nabla \chi = 0
\nonumber \\
& & \frac{\partial \eta}{\partial t} + (\vec \nabla \nu + \alpha \vec \nabla \chi + \beta \vec \nabla \eta + \sigma \vec \nabla s)
\cdot \vec \nabla \eta = 0
\nonumber \\
& & \frac{\partial s}{\partial t} + (\vec \nabla \nu + \alpha \vec \nabla \chi + \beta \vec \nabla \eta + \sigma \vec \nabla s)
\cdot \vec \nabla s = 0
\label{lagmul4}
\enr
Those are algebraic equations for $\alpha, \beta, \sigma$, which can be solved such that $\alpha, \beta, \sigma$ can be written
as functionals of $\chi,\eta,\nu,s$, resulting eventually in the description of non-barotropic magnetohydrodynamics
in terms of five functions: $\nu,\rho,\chi,\eta,s$.
Let us introduce the notation:

\beq
\alpha_i \equiv (\alpha, \beta, \sigma),  \chi_i\equiv (\chi,\eta,s),
 k_i \equiv -\frac{\partial \chi_i}{\partial t} - \vec \nabla \nu \cdot \vec \nabla \chi_i
 \label{ali}
\enq
$ i\in(1,2,3)$. In terms of the above notation \ern{lagmul4} takes the form:

\beq
k_i =\alpha_j \vec \nabla \chi_i \cdot \vec \nabla \chi_j, \qquad j\in(1,2,3)
\label{kieq}
\enq
in which the Einstein summation convention is assumed. Let us define the matrix:

\beq
A_{ij} \equiv  \vec \nabla \chi_i \cdot \vec \nabla \chi_j
\label{Adef}
\enq
obviously this matrix is symmetric since $A_{ij}=A_{ji}$. Hence \er{kieq} takes the form:

\beq
k_i = A_{ij} \alpha_j, \qquad j\in(1,2,3)
\label{kieq2}
\enq
 Provided that the matrix $A_{ij}$ is not singular it has an inverse $A^{-1}_{ij}$ which can be written as:

\ber
A^{-1}_{ij}&=&\left|A\right|^{-1}\cdot
\nonumber \\
& & \hspace{-2cm}
\left(
\begin{array}{ccc}
 A_{22} A_{33}-A_{23}^2 & A_{13} A_{23}-A_{12} A_{33} & A_{12} A_{23}-A_{13} A_{22} \\
 A_{13} A_{23}-A_{12} A_{33} & A_{11} A_{33}-A_{13}^2 & A_{12} A_{13}-A_{11} A_{23} \\
 A_{12} A_{23}-A_{13} A_{22} & A_{12} A_{13}-A_{11} A_{23} & A_{11} A_{22}-A_{12}^2
\end{array}
\right)
\label{invAdef}
\enr
In which the determinant $\left|A\right|$ is given by the following equation:

\ber
\left|A\right| &=&
A_{11} A_{22} A_{33}-A_{11} A_{23}^2-A_{22} A_{13}^2
\nonumber \\
&-&A_{33} A_{12}^2 +2 A_{12} A_{13} A_{23}
\label{Adet}
\enr
In terms of the above equations the $\alpha_i$'s can be calculated as functionals of $\chi_i,\nu$ as
follows:

\beq
\alpha_i [\chi_i,\nu]= A^{-1}_{ij} k_j.
\label{aleq}
\enq
The velocity \ern{vform} can now be written as:

\ber
\vec v &=& \vec \nabla \nu + \alpha_i \vec \nabla \chi_i=  \vec \nabla \nu + A^{-1}_{ij} k_j \vec \nabla \chi_i
\nonumber \\
&=& \vec \nabla \nu - A^{-1}_{ij}\vec \nabla \chi_i (\frac{\partial \chi_j}{\partial t} + \vec \nabla \nu \cdot \vec \nabla \chi_j).
\label{vform2}
\enr
Provided that the $\chi_i$ is a coordinate basis in three dimensions, we may write:

\beq
\vec \nabla \nu= \vec \nabla \chi_n \frac{\partial \nu}{\partial \chi_n}, \qquad n\in(1,2,3).
\label{nudecom}
\enq
Inserting \ern{nudecom} into \ern{vform2} we obtain:

\beq
\vec v = - A^{-1}_{ij}\vec \nabla \chi_i \frac{\partial \chi_j}{\partial t}
\label{vform3}
\enq
in the above $\delta_{in}$ is a Kronecker delta. Thus the velocity $\vec v [\chi_i]$ is a functional of
$\chi_i$ only and is independent of $\nu$.

\subsection{Lagrangian Density and Variational Analysis}

Let us now rewrite the Lagrangian density $\hat {\cal L}[\chi_i,\nu,\rho]$ given in
\ern{Lagactionsimp6} in terms of the new variables:

\ber
 &  & \hat {\cal L}[\chi_i,\nu,\rho] = -\rho [\frac{\partial{\nu}}{\partial t} + \alpha_k [\chi_i,\nu] \frac{\partial{\chi_k}}{\partial t}
\nonumber \\
 &+& \  \varepsilon (\rho,\chi_3) + \frac{1}{2} \vec v [\chi_i]^2 ]
-\frac{1}{8 \pi}(\vec \nabla \chi_1 \times \vec \nabla \chi_2)^2
\label{Lagactionsimp7}
\enr
Let us calculate the variational derivative of $\hat {\cal L}[\chi_i,\nu,\rho]$ with respect to $\chi_i$ this will result in:

\ber
\delta_{\chi_i}\hat {\cal L} &\hspace{-0.4cm}=&\hspace{-0.4cm} -\rho [  \delta_{\chi_i} \alpha_k  \frac{\partial{\chi_k}}{\partial t} +
\alpha_{\underline{i}}  \frac{\partial \delta \chi_{\underline{i}}}{\partial t}
 +\  \delta_{\chi_i} \varepsilon (\rho,\chi_3) +  \delta_{\chi_i}\vec v \cdot \vec v ]
 \nonumber \\
&-& \frac{ \vec B} {4 \pi} \cdot  \delta_{\chi_i} (\vec \nabla \chi_1 \times \vec \nabla \chi_2)
\label{delchiLag}
\enr
in which the summation convention is not applied if the index is underlined.
However, due to \ern{vform2} we may write:

\beq
  \delta_{\chi_i}\vec v= \delta_{\chi_i} \alpha_k \vec \nabla \chi_k +   \alpha_{\underline{i}} \vec \nabla \delta \chi_{\underline{i}}.
\label{delchiv}
\enq
Inserting \ern{delchiv} into \ern{delchiLag} and rearranging the terms we obtain:

\ber
 \delta_{\chi_i}\hat {\cal L} &=& -\rho [  \delta_{\chi_i} \alpha_k  (\frac{\partial{\chi_k}}{\partial t}
+ \vec v \cdot \vec \nabla \chi_k )+
\alpha_{\underline{i}}  (\frac{\partial \delta \chi_{\underline{i}}}{\partial t}
\nonumber \\
&+& \vec v \cdot \vec \nabla \delta \chi_{\underline{i}}) +\  \delta_{\chi_i} \varepsilon (\rho,\chi_3) ]
 \nonumber \\
&-& \frac{ \vec B} {4 \pi} \cdot  \delta_{\chi_i} (\vec \nabla \chi_1 \times \vec \nabla \chi_2).
\label{delchiLag2}
\enr
Now by construction $\vec v$ satisfies \ern{lagmul4} and hence $\frac{\partial{\chi_k}}{\partial t}
+ \vec v \cdot \vec \nabla \chi_k  = 0$, this leads to:

\beq
 \delta_{\chi_i}\hat {\cal L} = -\rho \left[  \alpha_{\underline{i}} \frac{d \delta \chi_{\underline{i}}}{d t}
  + \delta_{\chi_i} \varepsilon (\rho,\chi_3) \right] - \frac{ \vec B} {4 \pi} \cdot  \delta_{\chi_i} (\vec \nabla \chi_1 \times \vec \nabla \chi_2).
\label{delchiLag3}
\enq
From now on the derivation proceeds as in \eqs (\ref{delActionchi},\ref{delActioneta},\ref{delActions}) resulting in \eqs
(\ref{albetaeq},\ref{sigmaeq}) and will not be repeated. The difference is that now $\alpha, \beta$ and $\sigma$ are
 not independent quantities, rather they depend through \ern{aleq}
on the derivatives of $\chi_i,\nu$. Thus, \eqs (\ref{delActionchi},\ref{delActioneta},\ref{delActions})
 are not first order equations in time but are second order equations. Now let us calculate the variational derivative
 with respect to $\nu$ this will result in the expression:

\beq
 \delta_{\nu} \hat {\cal L} = -\rho [ \frac{\partial{\delta \nu}}{\partial t} + \delta_{\nu} \alpha_n  \frac{\partial{\chi_n}}{\partial t}]
\label{delnuLag}
\enq
However, $\delta_{\nu} \alpha_k$ can be calculated from \ern{aleq}:

\beq
\delta_{\nu} \alpha_n = A^{-1}_{nj} \delta_{\nu} k_j = - A^{-1}_{nj} \vec \nabla \delta \nu \cdot \vec \nabla \chi_j
\label{delnualeq}
\enq
Inserting the above equation into \ern{delnuLag}:

\ber
 \delta_{\nu} \hat {\cal L} &=& -\rho [ \frac{\partial{\delta \nu}}{\partial t} - A^{-1}_{nj}  \vec \nabla \chi_j
  \frac{\partial{\chi_n}}{\partial t} \cdot \vec \nabla \delta \nu ] =
  \nonumber \\
  &-& \rho [ \frac{\partial{\delta \nu}}{\partial t} +  \vec v \cdot \vec \nabla \delta \nu ]=
   -\rho \frac{d{\delta \nu}}{d t}
\label{delnuLag2}
\enr
The above equation can be put to the form:

\beq
 \delta_{\nu} \hat {\cal L} = \delta \nu [\frac{\partial{\rho}}{\partial t} + \vec \nabla \cdot (\rho \vec v )]
-\frac{\partial{(\rho \delta \nu)}}{\partial t}- \vec \nabla \cdot (\rho \vec v \delta \nu )
\label{delnuLag3}
\enq
This obviously leads to the continuity \ern{masscon} and some boundary terms in space and time. The variational
derivative with respect to $\rho$ is trivial and the analysis is identical to the one in \ern{delActionrho} leading
to \ern{nueq}. To conclude this subsection let us summarize the equations of non-barotropic magnetohydrodynamics:

\ber
\frac{d \nu}{d t} &=& \frac{1}{2} \vec v^2 - w,
\frac{\partial{\rho}}{\partial t} + \vec \nabla \cdot (\rho \vec v ) = 0,
\nonumber \\
\frac{d \sigma}{dt} &=& T,
\frac{d \alpha}{dt} = \frac{\vec \nabla \eta \cdot \vec J}{\rho},
\frac{d \beta}{dt} = -\frac{\vec \nabla \chi \cdot \vec J}{\rho}
\label{equa}
\enr
in which $\alpha,\beta,\sigma,\vec v$ are functionals of $\chi,\eta,s,\nu$ as described above.
It is easy to show as in \ern{dvform2} that those variational equations are equivalent to the physical equations.

\subsection{Lagrangian Density in Explicit Form}

Let us put the Lagrangian density of \er{Lagactionsimp7}in a slightly more explicit form. First us look at the
term $\vec v^2$:

\ber
\vec v^2 &\hspace{-0.4cm}=& \hspace{-0.4cm}
 A^{-1}_{ij}\vec \nabla \chi_i \frac{\partial \chi_j}{\partial t} A^{-1}_{mn}\vec \nabla \chi_m \frac{\partial \chi_n}{\partial t}
\nonumber \\
 &=& A^{-1}_{ij} A^{-1}_{mn} A_{im} \frac{\partial \chi_j}{\partial t}  \frac{\partial \chi_n}{\partial t}
= A^{-1}_{jn} \frac{\partial \chi_j}{\partial t}  \frac{\partial \chi_n}{\partial t}
\label{vsq}
\enr
in the above we use \ern{vform3} and \ern{Adef}. Next let us look at the expression:

\ber
\alpha_k [\chi_i,\nu] \frac{\partial{\chi_k}}{\partial t}&\hspace{-0.7cm} =& \hspace{-0.7cm} A^{-1}_{kj} k_j \frac{\partial{\chi_k}}{\partial t}
 =-(\frac{\partial \chi_j}{\partial t} + \vec \nabla \nu \cdot \vec \nabla \chi_j)A^{-1}_{kj} \frac{\partial{\chi_k}}{\partial t}
 \nonumber \\
 &=& -A^{-1}_{jk} \frac{\partial \chi_j}{\partial t}  \frac{\partial \chi_k}{\partial t}
 - \frac{\partial \nu}{\partial \chi_m} \frac{\partial{\chi_m}}{\partial t}
\label{alpterm}
\enr
Inserting \er{vsq} and \er{alpterm} into \er{Lagactionsimp7} leads to a Lagrangian density of a  more
standard quadratic form:
\ber
 \hat {\cal L}[\chi_i,\nu,\rho] &=& \rho [\frac{1}{2} A^{-1}_{jn} \frac{\partial \chi_j}{\partial t}  \frac{\partial \chi_n}{\partial t}
+\frac{\partial \nu}{\partial \chi_m} \frac{\partial{\chi_m}}{\partial t}-
 \frac{\partial{\nu}}{\partial t}
 \nonumber \\
 &-&  \varepsilon (\rho,\chi_3)]
-\frac{1}{8 \pi}(\vec \nabla \chi_1 \times \vec \nabla \chi_2)^2.
\label{Lagactionsimp8}
\enr
We now define the metric $g_{jn} = A^{-1}_{jn}$ and obtain the geometrical Lagrangian:
\ber
 \hat {\cal L}[\chi_i,\nu,\rho] &=& \rho [\frac{1}{2} g_{jn} \frac{\partial \chi_j}{\partial t}  \frac{\partial \chi_n}{\partial t}
+\frac{\partial \nu}{\partial \chi_m} \frac{\partial{\chi_m}}{\partial t}-
 \frac{\partial{\nu}}{\partial t}
 \nonumber \\
 &-&  \varepsilon (\rho,\chi_3)]
-\frac{1}{8 \pi}(\vec \nabla \chi_1 \times \vec \nabla \chi_2)^2.
\label{Lagactionsimpg}
\enr
The Lagrangian is thus composed of a geometric kinetic term which is quadratic in the temporal
derivatives, a "gyroscopic" terms which is linear in the temporal derivative and a potential
 term which is independent of the temporal derivative.

\section {Noether Current}

Let us assume that all the equations of motion and boundary conditions of non barotropic MHD are satisfied. In this case
we have according to \ern{Lagactionsimp6}:
\beq
\delta A =  \int_{t_1}^{t_2} dt \int  d^3 x  \delta  \hat {\cal L}
= -\left. \int  d^3 x \rho \left[ \delta \nu + \alpha \delta \chi + \beta \delta \eta + \sigma \delta s \right] \right|_{t_1}^{t_2}.
 \label{NCLagactionsimp6}
\enq
For the current purpose it does not matter if $\alpha, \beta$ and $\sigma$ are independent variational variables or depend on
other variational variables through \ern{aleq}. Now suppose that the variations $\delta \nu,  \delta \chi, \delta \eta,  \delta s$ are
symmetry variations such that $\delta A = 0$. In that case one obtains a conserved Noether current:
\beq
\delta J = - \int  d^3 x \rho \left[ \delta \nu + \alpha \delta \chi + \beta \delta \eta + \sigma \delta s \right].
 \label{NC1}
\enq
As the variations in the specific entropy $s$ will generally vary the specific internal energy term in the lagrangian we do not
expect non trivial entropy symmetry transformation and the action will only be invariant for $\delta s = 0$, hence:
\beq
\delta J = - \int  d^3 x \rho \left[ \delta \nu + \alpha \delta \chi + \beta \delta \eta \right].
 \label{NC1a}
\enq

\subsection {Lagrangian and Eulerian variations}

The value of a function $f$ can be modified by evaluating it at a different point in space, the difference between the new and old values would be:
\beq
 f (\vec x + \vec \xi) - f (\vec x) = \vec \xi \cdot \vec \nabla f
\label{spva}
\enq
in which $\vec x$ is a coordinate vector and $\vec \xi$ is a displacement vector, the equality is correct to first order in $\vec \xi$.
 Alternatively we can modify the value of a function by changing it to a different function $f'$, in this case the difference between the new and old values would be:
 \beq
\delta f = f' (\vec x) - f (\vec x)
\label{del}
\enq
 for a small $\delta f$ this just the standard variation of variational analysis or an Eulerian variation.
 Finally we can do both, in the last case the difference between the new and old values would be:
\beq
\Delta f = f' (\vec x + \vec \xi) - f (\vec x)
\label{Del}
\enq
Hence:
\beq
\Delta f = f' (\vec x + \vec \xi) - f (\vec x + \vec \xi) + f (\vec x + \vec \xi) - f (\vec x).
\label{Del2}
\enq
Keeping only first order terms we obtain:
\beq
\Delta f = \delta f + \vec \xi \cdot \vec \nabla f \Rightarrow \delta f = \Delta f  -  \vec \xi \cdot \vec \nabla f.
\label{Del3}
\enq
in which $\Delta$ is a Lagrangian variation.

Now suppose that a specific function is connected to a fluid element in such a way
that its value in space is determined only by fluid element location. And suppose that the fluid element is displaced as
dictated by the flow. Such a function would be denoted a label of the flow and its material derivative would vanish. Moreover, for a label:
\beq
 f' (\vec x + \vec \xi) = f (\vec x) \Rightarrow  \Delta f = 0
\label{Del4}
\enq
in order to change the value of a label in a certain point in space the fluid element must be displaced and another (with a different label value) must take its place. If follows from \ern{Del3} that for a label:
\beq
\delta f = - \vec \xi \cdot \vec \nabla f.
\label{Del5}
\enq
Now suppose we have a set of three labels $\tilde \chi_i$ such that:
\beq
\delta \tilde \chi_i= - \vec \xi \cdot \vec \nabla \tilde \chi_i = -  \xi_k \frac{\partial \tilde \chi_i}{\partial x_k},
\label{Del5b}
\enq
in which we use the Einstein summation convention and $x_k$ are Cartesian coordinates. The inverse of the matrix $\frac{\partial \tilde \chi_i}{\partial x_k}$ is $\frac{\partial x_k}{\partial \tilde \chi_i}$ as:
\beq
\frac{\partial \tilde \chi_i}{\partial x_k}  \frac{\partial x_j}{\partial \tilde \chi_i}  = \delta_k^j,
\label{Del5c}
\enq
$\delta_k^j$ is Kronecker's delta. It thus follows that one can calculate the displacement vector $\vec \xi$ as follows:
\beq
 \xi_k = - \frac{\partial x_k}{\partial \tilde \chi_i} \delta \tilde \chi_i \Rightarrow
  \vec \xi = -\frac{\partial \vec r}{\partial \tilde \chi_i} \delta \tilde \chi_i
\label{Del5d}
\enq

\subsection {Noether current for label symmetries}

We now study the form of the Noether current \ern{NC1a} for the case of label symmetry transformations.
It is clear from \ern{lagmul} that $\chi,\eta$ can be taken to be labels. Hence we can write the conserved Noether current defined in \ern{NC1a} as:
\beq
\delta J = - \int  d^3 x \rho \left[ \Delta \nu  -  \vec \xi \cdot \vec \nabla \nu - \alpha \vec \xi \cdot \vec \nabla \chi - \beta \vec \xi \cdot \vec \nabla \eta  \right].
 \label{NC2}
\enq
Or using \ern{vform} as:
\beq
\delta J =  \int  d^3 x \rho \left[\vec \xi \cdot (\vec v - \sigma \vec \nabla s) - \Delta \nu  \right].
 \label{NC3}
\enq
We use the topological velocity defined in \ern{vt}:
\beq
\vec v_t = \vec \nabla \nu + \alpha \vec \nabla \chi + \beta \vec \nabla \eta
\label{vt2}
\enq
and write:
\beq
\delta J =  \int  d^3 x \rho \left[\vec \xi \cdot \vec v_t  - \Delta \nu  \right].
 \label{NC4}
\enq
 Suppose now that we are considering label symmetry transformations with the infinitesimal form: $\tilde \chi_i + \delta \tilde \chi_i$ this type of transformations will induce  a transformation on other
functions ($\nu$ as an example) which could be thought of as functions of labels a transformation of the form:
\beq
\delta \nu=  \nu (\tilde \chi_i + \delta \tilde \chi_i) - \nu (\tilde \chi_i ) = \delta \tilde \chi_i \partial_{\tilde \chi_i} \nu =
- \vec \xi \cdot \vec \nabla \tilde \chi_i \partial_{\tilde \chi_i} \nu = - \vec \xi \cdot \vec \nabla \nu.
 \label{Nutr}
\enq
Hence by \ern{Del3}:
\beq
\Delta \nu =  0.
 \label{Nutr2}
\enq
it follows that for an induced infinitesimal label transformation any function will transform as a label.
From \ern{Nutr2} it follows that the Noether current will take the following form for a symmetry label transformation:
\beq
\delta J =  \int  d^3 x \rho \vec \xi \cdot \vec v_t.
 \label{NC5}
\enq
This Noether current form is identical to equation (47) of \cite{metra} and equation (14) of \cite{chiettrans}, which were derive from a Lagrangian variational principle.  We note, however, that this form is limited to the case of label transformations and the general form given in \ern{NC1} allows us to exploit larger symmetry groups. Next we will study some symmetry transformations of the action $A$, in order to do this we shall first introduce the Load and Metage quantities.

\section{Load and Metage}
\label{inverse}

The following section follows closely a similar section in \cite{YaLy,simpvarYah,metra,anewdif}. Consider a thin tube surrounding a magnetic field line, the magnetic flux contained within the tube is:
\beq
\Delta \Phi =
\int \vec B \cdot d \vec S
\label{flux}
\enq
and the mass
contained with the tube is:
\beq
\Delta M = \int \rho d\vec l \cdot d \vec S,
\label{Mass}
\enq
in which $dl$ is a length element
along the tube. Since the magnetic field lines move with the flow
by virtue of \ern{Beq} and \ern{masscon} both the quantities $\Delta \Phi$ and
$\Delta M$ are conserved and since the tube is thin we may define
the conserved magnetic load:
\beq
\lambda = \frac{\Delta M}{\Delta
\Phi} = \oint \frac{\rho}{B}dl,
\label{Load}
\enq
in which the
above integral is performed along the field line. Obviously the
parts of the line which go out of the flow to regions in which
$\rho=0$ have a null contribution to the integral.
Notice that $\lambda$ is a {\bf single valued} function that can be measured in principle.
Since $\lambda$
is conserved it satisfies the equation:
\beq
 \frac{d \lambda }{d t} = 0.
\label{Loadcon}
\enq
This can be viewed as a manifestation of the frozen-in law of $\frac{B}{\rho}$.
By construction surfaces of constant magnetic load move with the flow and contain
magnetic field lines. Hence the gradient to such surfaces must be orthogonal to
the field line:
\beq
\vec \nabla \lambda \cdot \vec B = 0.
\label{Loadortho}
\enq
Now consider an arbitrary comoving point on the magnetic field line and denote it by $i$,
and consider an additional comoving point on the magnetic field line and denote it by $r$.
The integral:
\beq
\mu(r)  = \int_i^r \frac{\rho}{B}dl + \mu(i),
\label{metage}
\enq
is also a conserved quantity which we may denote following Lynden-Bell \& Katz \cite{LynanKatz}
as the magnetic metage. $\mu(i)$ is an arbitrary number which can be chosen differently for each
magnetic line. By construction:
\beq
 \frac{d \mu }{d t} = 0.
\label{metagecon}
\enq
This can be viewed as another manifestation of the frozen-in law of $\frac{B}{\rho}$.
Also it is easy to see that by differentiating along the magnetic field line we obtain:
\beq
 \vec \nabla \mu \cdot \vec B = \rho.
\label{metageeq}
\enq
Notice that $\mu$ will be generally a {\bf non single valued} function, we will show later in this paper
that symmetry to translations in $\mu$; will generate through the Noether theorem the conservation of the
magnetic cross helicity.

At this point we have two comoving coordinates of flow, namely $\lambda,\mu$ obviously in a
three dimensional flow we also have a third coordinate. However, before defining the third coordinate
we will find it useful to work not directly with $\lambda$ but with a function of $\lambda$.
Now consider the magnetic flux within a surface of constant load $\Phi(\lambda)$. The magnetic flux is a conserved quantity
and depends only on the load $\lambda$ of the surrounding surface. Now we define the quantity:
\beq
 \chi = \frac{\Phi(\lambda)}{2\pi}.
\label{chidef}
\enq
Obviously $\chi$ satisfies the equations:
\beq
\frac{d \chi}{d t} = 0, \qquad \vec B \cdot \vec \nabla \chi = 0.
\label{chieq}
\enq
Let us now define an additional comoving coordinate $\eta^{*}$
since $\vec \nabla \mu$ is not orthogonal to the $\vec B$ lines we can choose $\vec \nabla \eta^{*}$ to be
orthogonal to the $\vec B$ lines and not be in the direction of the $\vec \nabla \chi$ lines,
that is we choose $\eta^{*}$
not to depend only on $\chi$. Since both $\vec \nabla \eta^{*}$ and $\vec \nabla \chi$ are orthogonal to $\vec B$,
$\vec B$ must take the form:
\beq
\vec B = A \vec \nabla \chi \times \vec \nabla \eta^{*}.
\enq
However, using \ern{Bcon} we have:
\beq
\vec \nabla \cdot \vec B = \vec \nabla A \cdot (\vec \nabla \chi \times \vec \nabla \eta^{*})=0.
\enq
Which implies that $A$ is a function of $\chi,\eta^{*}$. Now we can define a new comoving function
$\eta$ such that:
\beq
\eta = \int_0^{\eta^{*}}A(\chi,\eta^{'*})d\eta^{'*}, \qquad \frac{d \eta}{d t} = 0.
\enq
In terms of this function we obtain the Sakurai (Euler potentials) presentation:
\beq
\vec B = \vec \nabla \chi \times \vec \nabla \eta.
\label{Bsakurai}
\enq
And the density is now given by the Jacobian:
\beq
\rho = \vec \nabla \mu \cdot (\vec \nabla \chi \times \vec \nabla \eta)
=\frac{\partial(\chi,\eta,\mu)}{\partial(x,y,z)}.
\label{metagecon2}
\enq
It can easily be shown using the fact that the labels are comoving that the above forms of
$\vec B$ and $\rho$ satisfy \ern{Beq}, \ern{Bcon} and \ern{masscon} automatically.

We can now write a Lagrangian density in terms of the labels, in which $\rho$ is no longer an independent variational
variable but rather a quantity dependent on $\mu$ through \ern{metagecon2}. The Lagrangian density of \ern{Lagactionsimp6} takes the form:
\ber
& & \hspace{-1.2cm}\hat {\cal L}[\alpha,\beta,\chi,\eta,\mu,\nu,\sigma,s] = -\frac{\partial(\chi,\eta,\mu)}{\partial(x,y,z)} \left[\frac{\partial{\nu}}{\partial t} + \alpha \frac{\partial{\chi}}{\partial t} + \beta \frac{\partial{\eta}}{\partial t} + \sigma \frac{\partial{s}}{\partial t}
+  \varepsilon (\frac{\partial(\chi,\eta,\mu)}{\partial(x,y,z)},s) \right.
\nonumber \\
&+&  \left. \frac{1}{2} (\vec \nabla \nu + \alpha \vec \nabla \chi +  \beta \vec \nabla \eta +  \sigma \vec \nabla s )^2 \right]
-\frac{1}{8 \pi}(\vec \nabla \chi \times \vec \nabla \eta)^2
\label{Lagactionsimp6bb}
\enr

Notice however, that $\eta$ is defined in a non unique way since one can redefine
$\eta$ for example by performing the following transformation: $\eta \rightarrow \eta + f(\chi)$
in which $f(\chi)$ is an arbitrary function.
The comoving coordinates $\chi,\eta$ serve as labels of the magnetic field lines.
Moreover the magnetic flux can be calculated as:
\beq
\Phi = \int \vec B \cdot d \vec S = \int d \chi d \eta.
\label{phichieta}
\enq
In the case that the surface integral is performed inside a load contour we
obtain:
\beq
\Phi (\lambda) = \int_{\lambda} d \chi d \eta= \chi \int_{\lambda} d \eta =\left\{
\begin{array}{c}
 \chi [\eta] \\
  \chi (\eta_{max}-\eta_{min}) \\
\end{array}
\right.
\label{Phila}
\enq
 There are two cases involved; in one case the load surfaces are topological cylinders;
 in this case $\eta$ is not single valued and hence we obtain the upper value for $\Phi (\lambda)$.
 In a second case the load surfaces are topological spheres; in this case $\eta$ is single valued
 and has minimal $\eta_{min}$ and maximal $\eta_{max}$  values. Hence the lower value of $\Phi (\lambda)$ is obtained.
 For example in some cases $\eta$ is identical to twice the latitude angle $\theta$.
 In those cases $\eta_{min}=0$ (value at the "north pole") and $\eta_{max}= 2 \pi$
 (value at the "south pole").

Comparing the above \eq \ with \ern{chidef} we derive that $\eta$ can be either
{\bf single valued} or {\bf not single valued} and that
its discontinuity across its cut in the non single valued case is $[\eta] =2 \pi$.

The triplet $\chi,\eta,\mu$ will suffice to label any fluid element in three dimensions. But for a non-barotropic
flow there is also another possible label $s$ which is comoving according to \ern{Ent}. The question then arises of the relation of this
label to the previous three. As one needs to make a choice regarding the preferred set of labels it seems that the physical
ones are $\chi,\eta,s$ in which we use the surfaces on which the magnetic fields lie and the entropy, each label has an obvious
physical interpretation. In this case we must look at $\mu$ as a function of $\chi,\eta,s$. If the magnetic field lines lie on entropy
surface then $\mu$ regains its status as an independent label. The density can now be written as:
\beq
\rho = \frac{\partial \mu}{ \partial s} \frac{\partial(\chi,\eta,s)}{\partial(x,y,z)}.
\label{metagecon3}
\enq
Now as $\mu$ can be defined for each magnetic field line separately according to \ern{metage} it is obvious that
such a choice exist in which $\mu$ is a function of $s$ only. One may also think of the entropy $s$ as a functions
$\chi,\eta,\mu$. However, if one change $\mu$ in this case this generally entails a change in $s$ and the symmetry described in
\ern{metage} is lost in the Action. In what follows we shall ignore the status of $s$ as a label and consider it as a variational variable which only attains a status of a label at the variational extremum.

\section {The labelling symmetry group and its subgroups}

It is obvious that the choice of fluid labels is quite arbitrary. However, when enforcing the $\chi, \eta, \mu$ coordinate
system satisfying \ern{metagecon2} the choice is restricted to $\tilde \chi, \tilde \eta, \tilde \mu$ such that:
\beq
 {\partial (\tilde \chi, \tilde \eta, \tilde \mu) \over
 \partial (\chi, \eta, \mu)} = 1.
 \label{alphasym}
  \enq
Moreover the Euler potential magnetic field representation:
\beq
\vec B = \vec \nabla \chi \times \vec \nabla \eta,
\label{Bsakurai2b}
\enq
reduces the choice further to:
\beq
 {\partial (\tilde \chi, \tilde \eta) \over  \partial (\chi, \eta)} = 1.
 \label{alphasym2}
\enq
We further notice that in the Eulerian variation principle approach the label symmetry cannot be realized unless it is coupled
to transformation of other variational variables that is the label transformation induces transformation on $\alpha$ and $\beta$  as follows:
\ber
  \tilde \alpha &=&  \alpha \partial_{\tilde{\chi}} \chi + \beta \partial_{\tilde{\chi}} \eta
 \nonumber \\
 \tilde \beta &=&  \alpha \partial_{\tilde{\eta}} \chi + \beta \partial_{\tilde{\eta}} \eta
 \label{alphasym3}
\enr
From \ern{Lagactionsimp6bb} it follows that:
\beq
\hat {\cal L}[\alpha,\beta,\chi,\eta,\mu,\nu,\sigma,s] - \hat {\cal L}[\tilde \alpha, \tilde \beta,\tilde \chi, \tilde \eta, \tilde \mu,\nu,\sigma,s] = 0
\label{Lagactionsimp6bb2}
\enq
hence the label transformation is a symmetry transformation. Now suppose that we consider $\nu$ as a function of the labels:
\beq
\nu(x,y,z,t) = \bar \nu(\chi,\eta,\mu,t)
\label{nulab}
\enq
in that case replacing  $(\chi,\eta,\mu) \rightarrow (\tilde \chi, \tilde \eta, \tilde \mu)$ in the above equation will yield
a different function $\tilde \nu$ of the coordinates such that:
\beq
 \bar \nu(\tilde \chi, \tilde \eta, \tilde \mu,t) = \tilde \nu(x,y,z,t)
\label{nulab2}
\enq
From this point of view which we adopt in the current paper, symmetry can be only achieved if:
\beq
\int d^3 x \left[\hat {\cal L}[\alpha,\beta,\chi,\eta,\mu,\bar \nu(\chi,\eta,\mu,t),\sigma,s] -
 \hat {\cal L}[\tilde \alpha, \tilde \beta,\tilde \chi, \tilde \eta, \tilde \mu,\bar \nu(\tilde \chi, \tilde \eta, \tilde \mu,t),\sigma,s]\right] = 0
\label{Lagactionsimp6bb3}
\enq

\subsection{Metage translations}

In what follows we consider the transformation (see also \ern{metage}):
\beq
\tilde \chi = \chi, \tilde \eta = \eta, \tilde \mu = \mu + a (\chi,\eta)
\enq
Hence $a$ is a label displacement which may be different for each magnetic field line, as the field line
is closed one need not worry about edge difficulties. This transformation satisfies trivially the conditions
(\ref{alphasym},\ref{alphasym2}).
If we take the infinitesimal symmetry transformation $\delta \mu = a, \delta \chi =\delta \eta =0$ we can calculate the associated fluid element displacement with this relabelling using \ern{Del5d} and \ern{metageeq}.
\beq
 \vec \xi = -{\partial \vec r \over \partial \mu}\delta \mu = -\delta \mu \frac{\vec B}{\rho}.
\label{ximu}
 \enq
Inserting \ern{ximu} into \ern{NC5} we obtain the conservation law:
\beq
\delta J =\int d^3 x \rho \vec v_t  \cdot \vec \xi = -\int d^3 x \delta \mu \vec v_t  \cdot \vec B
\label{Noether2}
\enq
In the simplest case we may take $\delta \mu$ to be a small constant, hence:
\beq
\delta J = - \delta \mu \int d^3 x \vec v_t \cdot \vec B = - \delta \mu H_{CNB}
\label{Noether3}
\enq
Where $H_{CNB}$ is the non barotropic global cross helicity \cite{Webb1,Yahalomhel2,Yahalomhel3} defined as:
\beq
H_{CNB} \equiv  \int d^3 x \vec v_t \cdot \vec B.
\label{Noether4}
\enq
 We thus obtain
the conservation of non-barotropic cross helicity using the Noether theorem and the symmetry group of metage translations.
Of course one can perform a different translation on each magnetic field line, in this case one obtains:
\beq
\delta J =  -\int d^3 x \delta \mu \vec v_t  \cdot \vec B =
 -\int d \chi d \eta \delta \mu  \oint_{\chi,\eta} d\mu \rho^{-1}  \vec v_t  \cdot \vec B
\label{Noether5}
\enq
Now since $\delta \mu$ is an arbitrary (small) function of $\chi,\eta$ it follows that:
\beq
I = \oint_{\chi,\eta} d\mu \rho^{-1}  \vec v_t  \cdot \vec B
\label{Noether6}
\enq
is a conserved quantity for each magnetic field line. Along a magnetic field line the following equations hold:
\beq
d\mu = \vec \nabla \mu \cdot d \vec r = \vec \nabla \mu \cdot \hat{B} dr = \frac{\rho}{B} dr
\label{Noether7}
\enq
in the above $\hat{B}$ is an unit vector in the magnetic field direction an \ern{metageeq} is used.
Inserting \ern{Noether7} into \ern{Noether6} we obtain:
\beq
I = \oint_{\chi,\eta} dr  \vec v_t  \cdot \hat B = \oint_{\chi,\eta} d \vec r \cdot  \vec v_t.
\label{Noether8}
\enq
which is just the circulation of the topological velocity along the magnetic field lines. This quantity can be
written in terms of the generalized Clebsch representation of the velocity \ern{vform} as:
\beq
I  = \oint_{\chi,\eta} d \vec r \cdot  \vec v_t = \oint_{\chi,\eta} d \vec r \cdot  \vec \nabla \nu = [\nu].
\label{Noether9}
\enq
$[\nu]$ is the discontinuity of $\nu$. This was shown to be equal to the amount of non barotropic cross helicity per unit
 of magnetic flux in \ern{loc_} \cite{Yahalomhel2,Yahalomhel3}.
\begin{equation}
I=[\nu]= \frac{dH_{CNB}}{d \Phi}.
\label{loc_2}
\end{equation}

\subsection{Transformations of magnetic surfaces}

Consider the following transformations:
\beq
 \tilde{\eta} = \eta + \delta \eta(\chi,\eta), \qquad \tilde{\chi} =  \chi + \delta \chi(\chi,\eta), \qquad \tilde{\mu} = \mu
\enq
in which $\delta \eta,\delta \chi$ are considered small in some sense. Inserting the above quantities into
\ern{alphasym2} and keeping only first order terms we arrive at:
\beq
\partial_{\eta} \delta \eta + \partial_{\chi} \delta \chi = 0.
\label{delchieteq}
\enq
This equation can be solved as follows:
\beq
 \delta \eta = \partial_{\chi} \delta f, \qquad  \delta \chi = -\partial_{\eta} \delta f,
 \label{delf}
\enq
in which $\delta f= \delta f (\chi,\eta)$ is an arbitrary small function. In this case we obtain a particle displacements of the
form:
\ber
 \vec \xi &=& -{\partial \vec r \over \partial \chi}\delta \chi -{\partial \vec r \over \partial \eta}\delta \eta =
 -\frac{1}{\rho} \left( \vec \nabla \eta \times \vec \nabla \mu \ \delta \chi +  \vec \nabla \mu \times \vec \nabla \chi \ \delta \eta  \right)
 \nonumber \\
 &=& \frac{\vec \nabla \mu}{\rho} \times ( \vec \nabla \eta \delta \chi - \vec \nabla \chi \delta \eta )
\label{xi2}
 \enr
A special case that satisfies \ern{delchieteq} is the case of a constant $\delta \chi$ and  $\delta \eta$, those two independent displacements
lead to two new topological conservation laws:
\ber
\delta J_\chi &=&  \delta \chi \int d^3 x \vec v _t \cdot \vec \nabla \mu \times \vec \nabla \eta= \delta \chi\  H_{CNB \chi},
\nonumber \\
\delta J_\eta &=& \delta \eta \int d^3 x \vec v _t \cdot \vec \nabla \chi \times \vec \nabla \mu = \delta \eta\  H_{CNB \eta}.
\label{Noether10}
\enr
Where the new non barotropic global cross helicities are defined as:
\beq
H_{CNB \chi} \equiv \int d^3 x \vec v _t \cdot \vec \nabla \mu \times \vec \nabla \eta, \quad
H_{CNB \eta} \equiv \int d^3 x \vec v _t \cdot \vec \nabla \chi \times \vec \nabla \mu
\label{Noether11}
\enq
We will find it useful to introduce the abstract "magnetic fields" as follows:
\beq
\vec B_\chi \equiv \vec \nabla \mu \times \vec \nabla \eta, \qquad \vec B_\eta \equiv \vec \nabla \chi \times \vec \nabla \mu
\label{AbstB}
\enq
In terms of which we obtain the new helicities in a more conventional form:
\beq
H_{CNB \chi} = \int d^3 x \vec v _t \cdot \vec B_\chi, \qquad
H_{CNB \eta} = \int d^3 x \vec v _t \cdot \vec B_\eta
\label{Noether12}
\enq
It is more plausible that those symmetries and conservation laws hold for magnetic field lines which lie on topological torii. In this case $\eta$ is non single valued \cite{YaLy} and thus the translation in this direction resembles moving fluid elements along closed loops. Both those helicities suffer a topological interpretation in terms  of the knottiness of the abstract magnetic field lines and the flow lines. Finally we remark that for barotropic MHD $\vec v_t$ can be replaced with $\vec v$.

\section{Direct Derivation}

Before continuing to discuss the possible applications of the topological constants of motion, we shall demonstrate that the generalized cross helicities are indeed constant without relying on the Noether theorem.

\subsection{Direct derivation of the constancy of non barotropic cross helicity}

Taking the temporal derivative of the non barotropic cross helicity given in \ern{Noether4} we obtain:
\beq
\frac{d H_{CNB}}{dt} = \int d^3 x \left[ \partial_t \vec v_t \cdot \vec B +  \vec v_t \cdot \partial_t \vec B \right].
\label{ddhcnb1}
\enq
in the above $\frac{d }{dt}$ is an ordinary temporal derivative, and we use the notation: $\partial_t \equiv \frac{\partial }{\partial t}$. Using
\ern{Beq} it follows that:
\beq
 \vec v_t \cdot \partial_t \vec B = \vec v_t \cdot \vec \nabla \times (\vec v \times \vec B)=
 \vec \nabla \cdot \left( (\vec v \times \vec B) \times  \vec v_t \right) + (\vec v \times \vec B) \cdot \vec \omega_t
\label{ddhcnb2}
\enq
in which we have used a standard identity of vector analysis and the definition:
\beq
  \vec \omega_t =  \vec \nabla \times \vec v_t = \vec \omega - \vec  \nabla \sigma \times \vec  \nabla s
\label{ddhcnb3}
\enq
$\vec v_t$ is defined in \ern{vt}. Next we calculate:
\beq
\partial_t \vec v_t \cdot \vec B = \vec B \cdot  \partial_t  \left(\vec v - \sigma \vec  \nabla s \right) = \vec B \cdot \left( \partial_t \vec v - \partial_t \sigma \vec  \nabla s
- \sigma \vec  \nabla  \partial_t s \right)
\label{ddhcnb4}
\enq
Taking into account Euler \ern{Euler} and the standard thermodynamic identities of \ern{thermodyn}:
\ber
\partial_t \vec v &=& -(\vec v \cdot \vec \nabla)\vec v  - \frac{1}{\rho}\vec \nabla p (\rho,s) +
\frac{(\vec \nabla \times \vec B) \times \vec B}{4 \pi \rho}
\nonumber \\
&=& \vec v \times \vec \omega - \vec \nabla (\frac{1}{2} \vec v^2)  - \vec \nabla w + T \vec \nabla s +
\frac{(\vec \nabla \times \vec B) \times \vec B}{4 \pi \rho}
\label{ddhcnb5}
\enr
Hence:
\beq
\vec B \cdot \partial_t \vec v =  \vec B \cdot \left(\vec v \times \vec \omega - \vec \nabla (\frac{1}{2} \vec v^2+w)   + T \vec \nabla s \right).
\label{ddhcnb6}
\enq
Taking into account \ern{sigmaeq} it follows that:
\beq
- \partial_t \sigma \vec  \nabla s = (\vec v \cdot \vec \nabla \sigma - T) \vec  \nabla s .
\label{ddhcnb7}
\enq
And taking into account \ern{Ent} it follows that:
\beq
- \sigma \vec  \nabla  \partial_t s = \sigma \vec  \nabla (\vec v \cdot \vec \nabla s).
\label{ddhcnb8}
\enq
Inserting \ern{ddhcnb6}, \ern{ddhcnb7} and \ern{ddhcnb8} into \ern{ddhcnb4} it follows that:
\beq
\partial_t \vec v_t \cdot \vec B  = \vec B \cdot \left( \vec v \times \vec \omega - \vec \nabla (\frac{1}{2} \vec v^2+w)
+ (\vec v \cdot \vec \nabla \sigma) \vec  \nabla s
+ \sigma \vec  \nabla (\vec v \cdot \vec \nabla s) \right)
\label{ddhcnb9}
\enq
Hence:
\ber
\partial_t \vec v_t \cdot \vec B&=&\vec B \cdot \left( \vec v \times \vec \omega + \vec \nabla \left(\sigma (\vec v \cdot \vec \nabla s)-\frac{1}{2} \vec v^2-w\right) \right.
\nonumber \\
&+& \left.(\vec v \cdot \vec \nabla \sigma) \vec  \nabla s - (\vec v \cdot \vec \nabla s) \vec  \nabla \sigma \right)
\nonumber \\
&& \hspace{-2cm} = \vec B \cdot \left( \vec v \times \vec \omega + \vec \nabla \left(\sigma (\vec v \cdot \vec \nabla s)-\frac{1}{2} \vec v^2-w\right)
+ (\vec \nabla \sigma  \times \vec  \nabla s)  \times \vec v \right)
\nonumber \\
& = & \vec B \cdot \left( \vec v \times \vec \omega_t + \vec \nabla \left(\sigma (\vec v \cdot \vec \nabla s)-\frac{1}{2} \vec v^2-w\right) \right)
\label{ddhcnb10}
\enr
Combining \ern{ddhcnb2} with \ern{ddhcnb10} we arrive at the result:
\ber
 \partial_t \vec v_t \cdot \vec B +  \vec v_t \cdot \partial_t \vec B &=& \vec \nabla \cdot \left( (\vec v \times \vec B) \times  \vec v_t \right)
 + \vec B \cdot \vec \nabla \left(\sigma (\vec v \cdot \vec \nabla s)-\frac{1}{2} \vec v^2-w\right)
 \nonumber \\
 &=& \vec \nabla \cdot \left[ (\vec v \times \vec B) \times  \vec v_t
 + \vec B  \left(\sigma (\vec v \cdot \vec \nabla s)-\frac{1}{2} \vec v^2-w\right)  \right]
\label{ddhcnb11}
\enr
in which we take into account \ern{Bcon}. Inserting \ern{ddhcnb11} into \ern{ddhcnb1} and using Gauss theorem we obtain a surface
integral:
\beq
\frac{d H_{CNB}}{dt} = \oint d \vec S \cdot \left[ (\vec v \times \vec B) \times  \vec v_t
 + \vec B  \left(\sigma (\vec v \cdot \vec \nabla s)-\frac{1}{2} \vec v^2-w\right)  \right].
\label{ddhcnb12}
\enq
The surface integral encapsulates the volume for which the non barotropic cross helicity is calculated. If the surface is taken at infinity
the magnetic fields vanish and thus:
\beq
\frac{d H_{CNB}}{dt} = 0
\label{ddhcnb13}
\enq
which means that $H_{CNB}$ is a constant of motion. We notice the complexity of the direct derivation with respect to the elegance and simplicity of the Noether theorem approach. However, obtaining the same result using different methods strengthens our confidence that no mathematical error was accidentally introduced.

\subsection{Direct derivation of the constancy of non barotropic $\chi$ cross helicity}

Taking the temporal derivative of the non barotropic $\chi$ cross helicity given in \ern{Noether12} we obtain:
\beq
\frac{d H_{CNB\chi}}{dt} = \int d^3 x \left[ \partial_t \vec v_t \cdot \vec B_\chi +  \vec v_t \cdot \partial_t \vec B_\chi \right].
\label{ddhcnbchi1}
\enq
Let us calculate $\partial_t \vec B_\chi$ where $\vec B_\chi$ is defined in \ern{AbstB}:
\beq
\vec B_\chi = \vec \nabla \mu \times \vec \nabla \eta.
\label{AbstBb}
\enq
It follows that:
\beq
\partial_t \vec B_\chi = \vec \nabla \partial_t \mu \times \vec \nabla \eta + \vec \nabla  \mu \times \vec \nabla \partial_t \eta.
\label{AbstBb2}
\enq
Using \ern{lagmul} and \ern{metagecon} we obtain:
\ber
\partial_t \vec B_\chi &=& \vec \nabla (-\vec v \cdot \vec \nabla \mu ) \times \vec \nabla \eta
+ \vec \nabla  \mu \times \vec \nabla (-\vec v \cdot \vec \nabla \eta)
\nonumber \\
 &=& \vec \nabla \times \left(\vec \nabla \mu (\vec v \cdot \vec \nabla \eta) - \vec \nabla \eta (\vec v \cdot \vec \nabla \mu)\right)
 \nonumber \\
 &=& \vec \nabla \times \left(\vec v  \times (\vec \nabla \mu \times \vec \nabla \eta) \right) =
 \vec \nabla \times \left(\vec v  \times  \vec B_\chi \right)
 \label{AbstBb3}
\enr
in which we used standard vector analysis identities. It thus follows that:
\beq
 \vec v_t \cdot \partial_t \vec B_\chi = \vec v_t \cdot \vec \nabla \times (\vec v \times \vec B_\chi)=
 \vec \nabla \cdot \left( (\vec v \times \vec B_\chi) \times  \vec v_t \right) + (\vec v \times \vec B_\chi) \cdot \vec \omega_t
\label{ddhcnbchi2}
\enq
Next we calculate:
\beq
\partial_t \vec v_t \cdot \vec B_\chi = \vec B_\chi \cdot  \partial_t  \left(\vec v - \sigma \vec  \nabla s \right)
= \vec B_\chi \cdot \left( \partial_t \vec v - \partial_t \sigma \vec  \nabla s - \sigma \vec  \nabla  \partial_t s \right)
\label{ddhcnbchi4}
\enq
Taking into account \ern{ddhcnb5}:
\beq
\vec B_\chi \cdot \partial_t \vec v =  \vec B_\chi \cdot \left(\vec v \times \vec \omega - \vec \nabla (\frac{1}{2} \vec v^2+w)   + T \vec \nabla s \right) + \vec B_\chi \cdot \frac{1}{\rho}\vec J \times \vec B
\label{ddhcnbchi6}
\enq
in which the current density is given by:
\beq
\vec J = \frac{\vec \nabla \times \vec B}{4 \pi} \quad \Rightarrow \vec \nabla \cdot \vec J = 0.
\label{J}
\enq
Now:
\beq
 \vec B_\chi \cdot \frac{1}{\rho}\vec J \times \vec B = \frac{1}{\rho} \vec J \cdot  \vec B \times \vec B_\chi.
\label{ddhcnbchi6a}
\enq
However:
\beq
\vec B \times \vec B_\chi = \vec B \times (\vec \nabla \mu \times \vec \nabla \eta) =
\vec \nabla \mu (\vec B \cdot \vec \nabla \eta) - \vec \nabla \eta (\vec B \cdot \vec \nabla \mu) = - \rho \vec \nabla \eta .
\label{ddhcnbchi6b}
\enq
It thus follows that:
\beq
 \vec B_\chi \cdot \frac{1}{\rho}\vec J \times \vec B = - \vec J \cdot \vec \nabla \eta = - \vec \nabla \cdot (\vec J \eta) .
\label{ddhcnbchi6c}
\enq
in which we used \ern{Bsakurai2} and \ern{metageeq}. Inserting \ern{ddhcnbchi6c} into \ern{ddhcnbchi6} will yield:
\beq
\vec B_\chi \cdot \partial_t \vec v =  \vec B_\chi \cdot \left(\vec v \times \vec \omega - \vec \nabla (\frac{1}{2} \vec v^2+w)   + T \vec \nabla s \right) - \vec \nabla \cdot (\vec J \eta)
\label{ddhcnbchi6d}
\enq
 Inserting \ern{ddhcnbchi6d}, \ern{ddhcnb7} and \ern{ddhcnb8} into \ern{ddhcnbchi4} it follows that:
\beq
\partial_t \vec v_t \cdot \vec B_\chi  = \vec B_\chi \cdot \left( \vec v \times \vec \omega - \vec \nabla (\frac{1}{2} \vec v^2+w)
+ (\vec v \cdot \vec \nabla \sigma) \vec  \nabla s
+ \sigma \vec  \nabla (\vec v \cdot \vec \nabla s) \right)- \vec \nabla \cdot (\vec J \eta)
\label{ddhcnbchi9}
\enq
Hence:
\beq
\partial_t \vec v_t \cdot \vec B_\chi
= \vec B_\chi \cdot \left( \vec v \times \vec \omega_t + \vec \nabla \left(\sigma (\vec v \cdot \vec \nabla s)-\frac{1}{2} \vec v^2-w\right) \right)- \vec \nabla \cdot (\vec J \eta)
\label{ddhcnbchi10}
\enq
Combining \ern{ddhcnbchi2} with \ern{ddhcnbchi10} we arrive at the result:
\ber
& & \partial_t \vec v_t \cdot \vec B_\chi +  \vec v_t \cdot \partial_t \vec B_\chi
 \nonumber \\
 &=& \vec \nabla \cdot \left( (\vec v \times \vec B_\chi) \times  \vec v_t - \vec J \eta \right)
 + \vec B_\chi \cdot \vec \nabla \left(\sigma (\vec v \cdot \vec \nabla s)-\frac{1}{2} \vec v^2-w\right)
 \nonumber \\
 &=& \vec \nabla \cdot \left[ (\vec v \times \vec B_\chi) \times  \vec v_t
 + \vec B_\chi  \left(\sigma (\vec v \cdot \vec \nabla s)-\frac{1}{2} \vec v^2-w\right) - \vec J \eta  \right]
\label{ddhcnbchi11}
\enr
in which we take into account \ern{AbstBb}. Inserting \ern{ddhcnbchi11} into \ern{ddhcnbchi1} and using Gauss theorem we obtain a surface
integral:
\ber
\frac{d H_{CNB \chi}}{dt} &=& \oint d \vec S \cdot \left[ (\vec v \times \vec B_\chi) \times  \vec v_t
 + \vec B_\chi  \left(\sigma (\vec v \cdot \vec \nabla s)-\frac{1}{2} \vec v^2-w\right)  - \vec J \eta  \right]
 \nonumber \\
 &-&  \int d \vec \Sigma \cdot \vec J [\eta] .
\label{ddhcnbchi12}
\enr
The surface integral encapsulates the volume for which the $\chi$ non barotropic cross helicity is calculated
and an additional surface integral is performed along the cut of $\eta$, in case that $\eta$ is not single valued (see \ern{Phila}). If the surface is taken at infinity the magnetic fields and current densities vanish and thus:
\beq
\frac{d H_{CNB \chi}}{dt} =  -  \int d \vec \Sigma \cdot \vec J [\eta]
\label{ddhcnbchi13}
\enq
hence for spherical topologies of magnetic field lines or for a current density $\vec J$ parallel to the cut we obtain:
\beq
\frac{d H_{CNB \chi}}{dt} = 0.
\label{ddhcnbchi14}
\enq
which means that $H_{CNB \chi}$ is a constant of motion. We notice the complexity of the direct derivation with respect to the elegance and simplicity of the Noether theorem approach. However, obtaining the same result using different methods strengthens our confidence that no mathematical error was accidentally introduced.

\subsection{Direct derivation of the constancy of non barotropic $\eta$ cross helicity}

Taking the temporal derivative of the non barotropic $\eta$ cross helicity given in \ern{Noether12} we obtain:
\beq
\frac{d H_{CNB\eta}}{dt} = \int d^3 x \left[ \partial_t \vec v_t \cdot \vec B_\eta +  \vec v_t \cdot \partial_t \vec B_\eta \right].
\label{ddhcnbeta1}
\enq
Let us calculate $\partial_t \vec B_\eta$ where $\vec B_\eta$ is defined in \ern{AbstB}:
\beq
\vec B_\eta = \vec \nabla \chi \times \vec \nabla \mu.
\label{AbstBbeta}
\enq
It follows that:
\beq
\partial_t \vec B_\eta = \vec \nabla \partial_t \chi \times \vec \nabla \mu + \vec \nabla  \chi \times \vec \nabla \partial_t \mu.
\label{AbstBb2eta}
\enq
Using \ern{lagmul} and \ern{metagecon} we obtain:
\ber
\partial_t \vec B_\eta &=& \vec \nabla (-\vec v \cdot \vec \nabla \chi ) \times \vec \nabla \mu
+ \vec \nabla  \chi \times \vec \nabla (-\vec v \cdot \vec \nabla \mu)
\nonumber \\
 &=& \vec \nabla \times \left(\vec \nabla \chi (\vec v \cdot \vec \nabla \mu) - \vec \nabla \mu (\vec v \cdot \vec \nabla \chi)\right)
 \nonumber \\
 &=& \vec \nabla \times \left(\vec v  \times (\vec \nabla \chi \times \vec \nabla \mu) \right) =
 \vec \nabla \times \left(\vec v  \times  \vec B_\eta \right)
 \label{AbstBb3eta}
\enr
in which we used standard vector analysis identities. It thus follows that:
\beq
 \vec v_t \cdot \partial_t \vec B_\eta = \vec v_t \cdot \vec \nabla \times (\vec v \times \vec B_\eta)=
 \vec \nabla \cdot \left( (\vec v \times \vec B_\eta) \times  \vec v_t \right) + (\vec v \times \vec B_\eta) \cdot \vec \omega_t
\label{ddhcnbeta2}
\enq
Next we calculate:
\beq
\partial_t \vec v_t \cdot \vec B_\eta = \vec B_\eta \cdot  \partial_t  \left(\vec v - \sigma \vec  \nabla s \right)
= \vec B_\eta \cdot \left( \partial_t \vec v - \partial_t \sigma \vec  \nabla s - \sigma \vec  \nabla  \partial_t s \right)
\label{ddhcnbeta4}
\enq
Taking into account \ern{ddhcnb5}:
\beq
\vec B_\eta \cdot \partial_t \vec v =  \vec B_\eta \cdot \left(\vec v \times \vec \omega - \vec \nabla (\frac{1}{2} \vec v^2+w)   + T \vec \nabla s \right) + \vec B_\eta \cdot \frac{1}{\rho}\vec J \times \vec B.
\label{ddhcnbeta6}
\enq
Now:
\beq
 \vec B_\eta \cdot \frac{1}{\rho}\vec J \times \vec B = \frac{1}{\rho} \vec J \cdot  \vec B \times \vec B_\eta.
\label{ddhcnbeta6a}
\enq
However:
\beq
\vec B \times \vec B_\eta = \vec B \times (\vec \nabla \chi \times \vec \nabla \mu) =
\vec \nabla \chi (\vec B \cdot \vec \nabla \mu) - \vec \nabla \mu (\vec B \cdot \vec \nabla \chi) = \rho \vec \nabla \chi.
\label{ddhcnbeta6b}
\enq
It thus follows that:
\beq
 \vec B_\eta \cdot \frac{1}{\rho}\vec J \times \vec B =  \vec J \cdot \vec \nabla \chi = \vec \nabla \cdot (\vec J \chi) .
\label{ddhcnbeta6c}
\enq
in which we used \ern{Bsakurai2} and \ern{metageeq}. Inserting \ern{ddhcnbeta6c} into \ern{ddhcnbeta6} will yield:
\beq
\vec B_\eta \cdot \partial_t \vec v =  \vec B_\eta \cdot \left(\vec v \times \vec \omega - \vec \nabla (\frac{1}{2} \vec v^2+w)   + T \vec \nabla s \right) + \vec \nabla \cdot (\vec J \chi)
\label{ddhcnbeta6d}
\enq
 Inserting \ern{ddhcnbeta6d}, \ern{ddhcnb7} and \ern{ddhcnb8} into \ern{ddhcnbeta4} it follows that:
\beq
\partial_t \vec v_t \cdot \vec B_\eta  = \vec B_\eta \cdot \left( \vec v \times \vec \omega - \vec \nabla (\frac{1}{2} \vec v^2+w)
+ (\vec v \cdot \vec \nabla \sigma) \vec  \nabla s
+ \sigma \vec  \nabla (\vec v \cdot \vec \nabla s) \right)+ \vec \nabla \cdot (\vec J \chi)
\label{ddhcnbeta9}
\enq
Hence:
\beq
\partial_t \vec v_t \cdot \vec B_\eta
= \vec B_\eta \cdot \left( \vec v \times \vec \omega_t + \vec \nabla \left(\sigma (\vec v \cdot \vec \nabla s)-\frac{1}{2} \vec v^2-w\right) \right)+ \vec \nabla \cdot (\vec J \chi)
\label{ddhcnbeta10}
\enq
Combining \ern{ddhcnbeta2} with \ern{ddhcnbeta10} we arrive at the result:
\ber
& & \partial_t \vec v_t \cdot \vec B_\eta +  \vec v_t \cdot \partial_t \vec B_\eta
 \nonumber \\
 &=& \vec \nabla \cdot \left( (\vec v \times \vec B_\eta) \times  \vec v_t +  \vec J \chi \right)
 + \vec B_\eta \cdot \vec \nabla \left(\sigma (\vec v \cdot \vec \nabla s)-\frac{1}{2} \vec v^2-w\right)
 \nonumber \\
 &=& \vec \nabla \cdot \left[ (\vec v \times \vec B_\eta) \times  \vec v_t
 + \vec B_\eta  \left(\sigma (\vec v \cdot \vec \nabla s)-\frac{1}{2} \vec v^2-w\right) +  \vec J \chi  \right]
\label{ddhcnbeta11}
\enr
in which we take into account \ern{AbstBbeta}. Inserting \ern{ddhcnbeta11} into \ern{ddhcnbeta1} and using Gauss theorem we obtain a surface
integral:
\beq
\frac{d H_{CNB \eta}}{dt} = \oint d \vec S \cdot \left[ (\vec v \times \vec B_\eta) \times  \vec v_t
 + \vec B_\eta  \left(\sigma (\vec v \cdot \vec \nabla s)-\frac{1}{2} \vec v^2-w\right)  +  \vec J \chi  \right].
\label{ddhcnbeta12}
\enq
The surface integral encapsulates the volume for which the $\eta$ non barotropic cross helicity is calculated. If the surface is taken at infinity the magnetic fields and current densities vanish and thus:
\beq
\frac{d H_{CNB \eta}}{dt} = 0.
\label{ddhcnbeta14}
\enq
which means that $H_{CNB \eta}$ is a constant of motion. We notice the complexity of the direct derivation with respect to the elegance and simplicity of the Noether theorem approach. However, obtaining the same result using different methods strengthens our confidence that no mathematical error was accidentally introduced.

\section{Possible Application}

 In his important review paper "Physics of magnetically confined plasmas" A. H. Boozer \cite{Boozer} states that: "A spiky
current profile causes a rapid dissipation of energy relative to magnetic helicity. If the evolution of a
magnetic field is rapid, then it must be at constant helicity." Usually topological conservation laws are used in order to deduce
lower bounds on the "energy" of the flow.
 Those bounds are only approximate in non ideal flows but due to their topological nature simulations show that
 they are approximately conserved even when the "energy" is not. For example it is easy to show that the "energy"
 is bounded from below by the non-barotropic cross helicity as follows (see \cite{anewdif}):
  \begin{equation}
H_{CNB} = \int  \vec{B}\cdot \vec{v_t}d^{3} x \leq \frac{1}{2}\int  \left( \vec{B}^2 + \vec{v_t}^2 \right) d^{3} x,
 \label{Globhelbound}
\end{equation}
\begin{equation}
H_{CNB} = \int  \vec{B}\cdot \vec{v_t}d^{3} x \leq \sqrt{\int  \vec{v_t}^2 d^{3} x}\sqrt{\int \vec{B}^2 d^{3} x},
 \label{Globhelbound2}
\end{equation}
the second equation is a result of the Cauchy-Schwartz inequality. In this sense a configuration with a highly complicated topology is more stable since its energy is bounded from below. It is a simple thing to show that similar bounds occur also for the $\chi$ and $\eta$ helicities:
 \begin{equation}
H_{CNB \chi} = \int  \vec{B_\chi}\cdot \vec{v_t}d^{3} x \leq \frac{1}{2}\int  \left( \vec{B_\chi}^2 + \vec{v_t}^2 \right) d^{3} x,
 \label{Globhelboundchi}
\end{equation}
\begin{equation}
H_{CNB \chi} = \int  \vec{B_\chi}\cdot \vec{v_t}d^{3} x \leq \sqrt{\int  \vec{v_t}^2 d^{3} x}\sqrt{\int \vec{B_\chi}^2 d^{3} x},
 \label{Globhelbound2chi}
\end{equation}
 \begin{equation}
H_{CNB \eta} = \int  \vec{B_\eta}\cdot \vec{v_t}d^{3} x \leq \frac{1}{2}\int  \left( \vec{B_\eta}^2 + \vec{v_t}^2 \right) d^{3} x,
 \label{Globhelboundeta}
\end{equation}
\begin{equation}
H_{CNB \eta} = \int  \vec{B_\eta}\cdot \vec{v_t}d^{3} x \leq \sqrt{\int  \vec{v_t}^2 d^{3} x}\sqrt{\int \vec{B_\eta}^2 d^{3} x},
 \label{Globhelbound2eta}
\end{equation}
Hence the kinetic energy is bounded by three differen bounds and so it the "total" energy. The importance of each of those bounds  is dependent on the flow.

\section{Conclusion}

We have derived a Noether current from an Eulerian variational principle on non-barotropic MHD, this was shown to lead to
to the conservation of non-barotropic cross helicity. The connection of the translation symmetry groups of labels to both
the global non barotropic cross helicity conservation law and the conservation law of circulations of topological velocity along
magnetic field lines was elucidated. The latter were shown to be equivalent to the amount of non barotropic cross helicity per unit
 of magnetic flux \cite{Yahalomhel2,Yahalomhel3,anewdif}. Further more we have shown that two additional cross helicity conservation laws exist the $\chi$ and $\eta$ cross helicities. Those lead to new bounds on  MHD  flows in addition to the bounds of the standard non-barotropic cross helicity discussed in \cite{Yahalomhel3} for ideal non-barotropic MHD. The importance of constants of motion for stability analysis is also discussed in \cite{Katz}. The significance of those constraints for non-ideal MHD and for plasma physics in general remains to be studied in future works.

It is shown that non-barotropic MHD can be derived from a variational principle of five functions.
The formalism is given in a Lagrangian presentation with a geometrical structure.

Possible applications include stability analysis of stationary MHD configurations and its possible utilization
for developing efficient numerical schemes for integrating the MHD equations.
It may be more efficient to incorporate the developed formalism in the framework of an existing
code instead of developing a new code from scratch. Possible existing codes are described
in \cite{Mignone,Narayan,Reisenegger}.
Applications of this study may be useful to both linear and non-linear stability analysis of known
 barotropic MHD configurations \cite{VMI,Kruskal,AHH,Katz,YahalomKatz,YahalomMonth}.
 As for designing efficient numerical schemes for integrating
the equations of fluid dynamics and MHD one may follow the
approach described in \cite{Yahalom,YahalomPinhasi,YahPinhasKop,OphirYahPinhasKop}.

Another possible application of the variational method is in deducing new analytic
solutions for the MHD equations. Although the equations are notoriously
difficult to solve being both partial differential equations and nonlinear, possible
solutions can be found in terms of variational variables. An example
for this approach is the self gravitating torus described in \cite{Yah3}.

One can use continuous symmetries which appear in the variational Lagrangian to derive new conservation laws
through the Noether theorem. An example for such derivation which
still lacks physical interpretation can be found in \cite{Yah5}. It may be that the
Lagrangian derived in \cite{Yah} has a larger symmetry group. And of course one anticipates
a different symmetry structure for the non-barotropic case.

Topological invariants have always been informative, and there are such invariants in MHD flows. For
example the two helicities  have long been useful in research into
the problem of hydrogen fusion, and in various astrophysical scenarios. In previous
works \cite{YaLy,Yah2,Yahalomhel} connections between helicities with symmetries of the barotropic fluid equations were made. The Noether current here derived may help us to identify and characterize as yet unknown topological invariants in MHD .

\vspace{0.5cm}
\noindent
{\LARGE \bf Acknowledgement}
\vspace{0.5cm}

\noindent This research was supported by the U.S. Department of Energy (DE-AC02-09CH11466).

\end {document}